\documentclass[twocolumn, superscriptaddress]{aastex63}
\usepackage{CJK}
\usepackage{url}
\usepackage{amsmath}
\usepackage[super]{nth}
\usepackage{microtype}
\usepackage{graphicx}

\received{August 20, 2021}
\revised{October 19, 2021}
\accepted{to AJ October 21, 2021}

\submitjournal{AAS Journals}

\begin{document}
\begin{CJK*}{UTF8}{gbsn}

\title{KELT-9 as an eclipsing double-lined spectroscopic binary: a unique and self-consistent solution to the system}
\shorttitle{KELT-9 b Dynamical Mass}
\shortauthors{Pai Asnodkar et al.}

\author[0000-0002-8823-8237]{Anusha Pai Asnodkar}
\affiliation{The Ohio State University, McPherson Laboratory, 140 W 18th Ave., Columbus, OH 43210, USA}
\author[0000-0002-4361-8885]{Ji Wang (王吉)}
\affiliation{The Ohio State University, McPherson Laboratory, 140 W 18th Ave., Columbus, OH 43210, USA}
\author{B. Scott Gaudi}
\affiliation{The Ohio State University, McPherson Laboratory, 140 W 18th Ave., Columbus, OH 43210, USA}
\author[0000-0001-9207-0564]{P. Wilson Cauley}
\affiliation{Laboratory for Atmospheric and Space Physics, University of Colorado Boulder, Boulder, CO 80303}
\author[0000-0003-3773-5142]{Jason D.\ Eastman}
\affiliation{Center for Astrophysics \textbar \ Harvard \& Smithsonian, 60 Garden St, Cambridge, MA 02138, USA}
\author{Ilya Ilyin}
\affiliation{Leibniz-Institut for Astrophysics Potsdam (AIP), An der Sternwarte 16, D–14482 Potsdam, Germany}
\author[0000-0002-6192-6494]{Klaus Strassmeier}
\affiliation{Leibniz-Institute for Astrophysics Potsdam (AIP), An der Sternwarte 16, D–14482 Potsdam, Germany}
\author{Thomas Beatty}
\affiliation{Department of Astronomy and Steward Observatory, University of Arizona, Tucson, AZ 85721}

\correspondingauthor{Anusha Pai Asnodkar}
\email{paiasnodkar.1@osu.edu}

\begin{abstract}
Transiting hot Jupiters present a unique opportunity to measure absolute planetary masses due to the magnitude of their radial velocity signals and known orbital inclination. Measuring planet mass is critical to understanding atmospheric dynamics and escape under extreme stellar irradiation. Here, we present the ultra-hot Jupiter system, KELT-9, as a double-lined spectroscopic binary. This allows us to directly and empirically constrain the mass of the star and its planetary companion, without reference to any theoretical stellar evolutionary models or empirical stellar scaling relations. Using data from the PEPSI, HARPS-N, and TRES spectrographs across multiple epochs, we apply least-squares deconvolution to measure out-of-transit stellar radial velocities. With the PEPSI and HARPS-N datasets, we measure in-transit planet radial velocities using transmission spectroscopy. By fitting the circular orbital solution that captures these Keplerian motions, we recover a planetary dynamical mass of $2.17 \pm 0.56\ \mathrm{M_J}$ and stellar dynamical mass of $2.11 \pm 0.78\ \mathrm{M_\odot}$, both of which agree with the discovery paper. Furthermore, we argue that this system, as well as systems like it, are highly overconstrained, providing multiple independent avenues for empirically cross-validating model-independent solutions to the system parameters. We also discuss the implications of this revised mass for studies of atmospheric escape.
\end{abstract}



\section{Introduction}\label{sec:intro}
\par Absolute mass is a critical but elusive quantity throughout the field of observational astronomy. Most empirical constraints on mass rely on the analysis of dynamical information of bodies interacting gravitationally, which are not always attainable from two-dimensional projections on the plane of the sky. Stellar eclipsing double-lined spectroscopic binaries (SB2s) are a classic case in which dynamical masses can be directly measured. Spectroscopic observations of two eclipsing stars, thus stars of known inclination, yield orbital velocities by harnessing the Doppler effect for light. Newton's fundamental law of gravitation can be combined with knowledge of the system's orbital motion to recover the dynamical masses of both stars purely empirically. These measurements calibrate the stellar evolution models that are typically used to determine the masses of stars and indirectly the planets they host \citep{Popper1980, Harmanec1988, Andersen1991, Torres2010, Stevens2018}. 

\par Mass is no less essential for the characterization of transiting planet systems. For example, determining the interior composition of terrestrial planets or measuring atmospheric escape on highly irradiated planets both require knowledge of the planet's mass. Transiting planets and their host stars are in an eclipsing orbital configuration that, in principle, enables a direct measurement of their masses. Transiting hot Jupiters (HJs), in particular, are particularly suitable for this task. Due to their relatively large mass and proximity to their host stars, HJs impart a stronger reflex motion through their gravitational influence on their host stars than other classes of planets. Thus they are the most accessible class of planets for transit and radial velocity (RV) techniques. The RV signature of the host star in principle can be constrained spectroscopically for the most massive and tightly bound HJs, even without high-precision radial velocity instruments. 

\par However, because of their extremely small planet/star flux ratios, most HJs systems are essentially eclipsing single-lined spectroscopic binaries (SB1s). Such systems do not allow for a unique solution to the masses and radii of the individual objects \citep{Seager2003}; rather there is a one-dimensional degeneracy between the mass and radius of the primary.  Thus any inferences about the mass of a transiting planet requires an external constraint on stellar mass, typically from stellar evolution models or empirical scaling relations between the properties of main-sequence stars \citep{Torres2010, Dotter2016, Choi2016, Duck2021}. However, the reliability of the estimates and uncertainties of these semi-empirical measurements has recently been called into question \citep{Tayar2020}. Systematic uncertainties in stellar properties that may deviate from the representative population used to calibrate evolutionary tracks can propagate into an incorrect estimation of planet mass and other planetary parameters. Furthermore, \citealt{Tayar2020} show that current planetary parameter uncertainties are often disproportionately underestimated relative to the uncertainties on the host star properties from which the planetary parameters are derived. Thus, obtaining masses by purely empirical methods is important because it provides a concrete check on the semi-empirical practices that are commonly employed.

\par If the radial velocity of transiting planet can be also measured, then the system essentially becomes an eclipsing SB2, and thus an empirical measurement of the system parameters, including the masses of the planet and star, can be made.  In this work, we apply the classical techniques used to analyze double-lined spectroscopic binaries to the KELT-9 system \citep{Gaudi2017}, using observations from the The Potsdam Echelle Polarimetric and Spectroscopic Instrument (PEPSI) spectrograph \citep{Strassmeier2015} on the Large Binocular Telescope and the HARPS-N spectrograph on Telescopio Nazionale Galileo (TNG). KELT-9 b is an ultra-hot Jupiter (UHJ) and the hottest known planet to-date ($T_{\mathrm{eq}} = 4050$ K). UHJs generally have orbital periods below 3 days, yielding orbital velocities in the range of hundreds of km s$^{-1}$; this is approximately 3 orders of magnitude greater than the orbital velocity of the host star. 

\par Furthermore, the planet's radial velocity can be measured from its atmospheric absorption signature observed during transit. \citealt{Snellen2010} is a pioneering work in this field, not only for providing the first measurement of winds on an exoplanet, but also for using this technique to empirically constrain the absolute masses of the planet and its host. CO was adopted as a tracer of HD 209458 b's absorption signature to simultaneously chart the planet's orbital motion and measure its day-to-nightside winds. As a result, the Doppler velocity shift of the planet's atmospheric absorption features over the course of a transit span a broader range of radial velocities (RVs) than that of the relatively stationary stellar spectral features. Transmission spectroscopy harnesses this distinction in velocity space to disentangle the planet's spectrum from its host star, allowing for a self-consistent measurement of the planet's absolute mass.

\par \citealt{Yan2018} adopt this method to measure the masses of KELT-9 b and its host ($M_{\mathrm{p}} = 3.23 \pm 0.94\ \mathrm{M_{\mathrm{J}}}$, $M_\star = 3.00 \pm 0.21\ \mathrm{M_\odot}$). However, they do not obtain an original measurement of $K_\star$, the stellar orbital velocity; instead they adopt the value given in \citealt{Gaudi2017} measured from TRES stellar spectra. In our work, we utilize the original TRES data in \citealt{Gaudi2017} as well as additional RVs from higher precision spectrographs, PEPSI (LBT) and HARPS-N (TNG), to constrain the dynamical masses of the KELT-9 system through a fully self-contained analysis. In combination with the Hipparcos parallax and SED fitting in \citealt{Gaudi2017} that allow us to constrain KELT-9 b's radius, these observations will provide a purely empirical measurement of the system parameters. 

\par The KELT-9 system is notably overconstrained. A key focus of this work is to emphasize that complementary observations (discussed in Section \ref{sec:overconstrained}) will contribute complete model-independent solutions to the system parameters. All of these constraints can be combined to determine the system parameters to higher accuracy. Improved precision of planetary parameters is critical for science cases pertinent to KELT-9 b, such as atmospheric escape. A tight constraint on mass will enable a more precise understanding of KELT-9 b's atmospheric escape since the mass loss rate measurement is dependent on planetary parameters, i.e. mass or surface gravity.

\par Like \citealt{Snellen2010}, we use a fully self-consistent technique analogous to the study of double-lined eclipsing binaries to obtain the absolute masses of the KELT-9 system. In section \ref{sec:observations}, we describe our new datasets from the PEPSI spectrograph as well as the archival HARPS-N observations we use in our analysis. In section \ref{sec:methods}, we outline the techniques we use to recover stellar and planetary orbital velocities and subsequently determine the absolute masses of the planet and star. In section \ref{sec:results}, we report our resulting mass constraints and compare our findings with previous literature. In section \ref{sec:discussion}, we discuss the ways in which the properties of the KELT-9 system are potentially empirically over-constrained.  We consider contributions from current and future complementary observations.  When combined, we argue that these will lead to a complete solution of system parameters. We also discuss the implications for atmospheric escape. Finally, we present our conclusions in \ref{sec:conclusion}.

\section{Observations} \label{sec:observations}


\begin{table}[t]
\caption{Datasets used to measure stellar RV: columns provide the name of the instrument, the date of the starting observation, the start and end times of observing, and the number of observations.}
\centering
    \begin{tabular}{lllll}
    \hline
    \hline
    Instrument & Night & $t_{\mathrm{start}}$ (UTC) & $t_{\mathrm{end}}$ (UTC) & $N_{\mathrm{obs}}$ \\ 
    \hline
    PEPSI & 2018-07-03 & 04:07:34.4 & 11:16:41.3 & 82 \\
    PEPSI & 2019-06-22 & 05:19:33.7 & 11:29:06.5 & 65 \\
    PEPSI & 2021-06-28 & 05:55:41.3 & 11:48:23.5 & 62$^\dagger$ \\
    HARPS-N & 2017-07-31 & 20:59:04.0 & 05:19:10.0 & 49 \\
    HARPS-N & 2018-07-20 & 21:20:24.0 & 05:09:58.0 & 46 \\ 
    TRES & 2014-2016 & & & 60$^\dagger$\\
    \hline
    \end{tabular}
    $^\dagger$Out-of-transit only.
\label{tab:datasets}
\end{table}

We observed two transits of KELT-9 b with the high-resolution \'echelle spectrograph PEPSI \citep{Strassmeier2015} on the Large Binocular Telescope (LBT; two 8.4-m mirrors, effective aperture of 11.8 m) in Arizona (see Table \ref{tab:datasets} for the specific nights of observations). PEPSI has a blue arm (nominally 3830--5440 {\AA}) and a red arm (nominally 5440--9070 {\AA}) with six cross-dispersers for full optical coverage. In this work, we use high-resolution data from the blue arm taken with cross-disperser 3 ($\sim$4750--5430 \text{\AA}, R=50,000) exclusively because of negligible telluric contamination \citep{Cauley2019}. The PEPSI pipeline produces wavelength-calibrated 1D spectra of each order which are then continuum-normalized, corrected for solar barycentric motion, and stitched into a single 1D spectral vector.

\par Our PEPSI dataset taken on 2018-07-03 (hereafter PEPSI 2018) was originally presented with an analysis of KELT-9 b's Balmer and metal lines in \citealt{Cauley2019}. In the blue arm, the spectra were taken with exposure times between 220 s and 387 s (depending on fluctuations in observing conditions) to approximately maintain a constant signal-to-noise ratio (SNR) of 210 in the continuum; in practice, the SNR ranged between 182 to 219 across observations. Additionally, we present a new dataset taken on 2019-06-22 (hereafter PEPSI 2019) for which observations began during the transit. The exposure times of the blue arm spectra were between 214 s and 270 s with continuum SNRs ranging between 286 and 321.

\par We utilized two archival HARPS-North (HARPS-N) datasets from the Italian Center for Astronomical Archives (IA2) Facility to increase the number of stellar RV samples for our measurement of the host star's orbital motion. HARPS-N is a high resolution (R$\sim$115,000) optical spectrograph (wavelength range between 3874 {\AA} and 6909 {\AA} ) on the Telescopio Nazionale Galileo (TNG) in La Palma, Spain. The first night of HARPS-North observations are from 2017-07-31 (hereafter HARPS-N 2017) and the other are from 2018-07-20 (hereafter HARPS-N 2018); both were originally presented in \citealt{Hoeijmakers2019}. The SNR of a given observation in these datasets ranges between ~35-140, depending on the order, at exposure times of 600 s. We retrieved the 1D, order-stitched spectra from the IA2 Archive Facility. 

\par Three of the datasets (PEPSI 2018,  HARPS-N 2017, HARPS-N 2018) include observations taken immediately before, during, and immediately after transit. PEPSI 2019 only includes observations during and immediately after transit. We converted all observation timings from their respective timing systems (PEPSI times are provided in both $\mathrm{JD_{UTC}}$ as well as $\mathrm{HJD_{UTC}}$; HARPS-N times are provided in $\mathrm{MJD_{UTC}}$) to $\mathrm{BJD_{TDB}}$ using the Time Utilities\footnote{\texttt{https://astroutils.astronomy.osu.edu/time/utc2bjd.html}} online software tool \citep{Eastman2012} to make them comparable with our revised ephemerides of the KELT-9 system  (see Table \ref{tab:system_params}). This is a crucial step for precision radial velocity measurements \citep{Eastman2010}, especially of atmospheric dynamics. The $\mathrm{JD_{UTC}}$ to $\mathrm{BJD_{TDB}}$ conversion (see \citealt{Eastman2010} for a detailed description of the difference between $\mathrm{JD_{UTC}}$ and $\mathrm{BJD_{TDB}}$) is not accounted for in previous literature, e.g. \citealt{Cauley2019}, and yields a difference up to 4.4 minutes in our PEPSI datasets. Note that the discovery paper \citep{Gaudi2017} ephemeris, which is commonly adopted in KELT-9 b literature, is also in the $\mathrm{BJD_{TDB}}$ timing system.

\begin{table*}[t]
\caption{KELT-9 system parameters: stellar and planetary parameters generally from the discovery paper as well as updated measurements of orbital configuration and ephemerides from this work.}
\centering
    \begin{tabular}{lllll}
    \hline
    \hline
    Parameter & Units & Symbol & Value & Source \\ 
    \hline
    Stellar parameters: \\
    \hspace{3mm} Stellar mass & $M_\odot$ & $M_\star$ & $2.52^{+0.25}_{-0.20}$ & \citealt{Gaudi2017} \\
    \hspace{3mm} Stellar radius & $R_\odot$ & $R_\star$ & $2.362^{+0.075}_{-0.063}$ & \citealt{Gaudi2017}\\
    \hspace{3mm} Stellar density & g $\mathrm{cm^{-3}}$ & $\rho_\star$ & $0.2702 \pm 0.0029$ & \citealt{Gaudi2017} \\
    \hspace{3mm} Effective temperature & K & $T_{\mathrm{eff}}$ & $10170 \pm 450$ & \citealt{Gaudi2017}\\
    \hspace{3mm} Projected rotational velocity & km s$^{-1}$ & $v\sin{i}$ & $111.4 \pm 1.3$ & \citealt{Gaudi2017} \\
    Planetary parameters: \\
    \hspace{3mm} Planet mass & $M_J$ & $m_{\mathrm{p}}$ & $2.88 \pm 0.84$ & \citealt{Gaudi2017} \\
    \hspace{3mm} Planet radius & $R_J$ & $R_{\mathrm{p}}$ & $1.891^{+0.061}_{-0.053}$ & \citealt{Gaudi2017} \\
    \hspace{3mm} Semi-major axis & AU & $a$ & $0.03462^{+0.00110}_{-0.00093}$ & \citealt{Gaudi2017} \\
    \hspace{3mm} Eccentricity & & $\varepsilon$ & $0$ & \citealt{Gaudi2017} \\
    \hspace{3mm} Spin-orbit alignment & degree & $\lambda$ & $-84.8\pm1.4$ & \citealt{Gaudi2017} \\
    \hspace{3mm} Orbital inclination & degree & $i_{\mathrm{orbit}}$ & $86.79\pm0.25$ & \citealt{Gaudi2017} \\
    Ephemeris: \\
    \hspace{3mm} Mid-transit time & $\mathrm{BJD_{TDB}}$ & $T_0$ & 2458566.436560 $\pm$ 0.000048 & This work \\
    \hspace{3mm} Time of secondary eclipse & $\mathrm{BJD_{TDB}}$ & $T_\mathrm{S}$ & 2458584.950546 $\pm$ 0.000048 & This work \\
    \hspace{3mm} Orbital period & days & $P$ & 1.48111890 $\pm$ 0.00000016 & This work \\
    \hspace{3mm} Ingress/egress transit duration & days & $\tau$ & $0.012808^{+0.000027}_{-0.000026}$ & This work \\
    \hspace{3mm} Total transit duration & days & $T_{14}$ & $0.15949\pm0.00011$ & This work \\
    \hline
    \end{tabular}
\label{tab:system_params}
\end{table*}

\section{Methods} \label{sec:methods}

\subsection{Stellar orbital properties}
To measure the dynamical mass of KELT-9 b, we need to know the orbital properties of the planet and its host star. We first recover the out-of-transit stellar velocities by least-squares deconvolution (LSD) of the stellar spectra across our datasets to fit for the stellar orbital velocity and systemic radial velocity as measured by each instrument.

\subsubsection{Least-squares deconvolution of stellar line profiles}
\label{sec:stellar_RV}
\par In subsequent analysis, $K_\star$ (stellar RV semi-amplitude) and $v_{\mathrm{sys},\ i}$ (systemic velocity measured by instrument $i$) are crucial values. The systemic radial velocity of the KELT-9 system has been a point of controversy in previous literature \citep{Gaudi2017, Hoeijmakers2019, Borsa2019}; see Table \ref{tab:revised_params} for a compilation of the different literature values. The standard procedure to recover stellar velocities is to centroid the cross-correlation function (CCF) profiles of the observations, where the CCF is the observed stellar spectra cross-correlated with a template spectrum corresponding to the star's effective temperature, with a Gaussian fit. This method becomes increasingly imprecise for fast rotators due to significant rotational broadening.

\begin{table}[b]
\centering
\caption{KELT-9 revised system parameters}
    \begin{tabular}{p{1.6cm}p{1cm}p{2.2cm}p{2.7cm}}
    \hline
    \hline
    Parameter & Units & Value & Source \\ 
    \hline
    $v_{\mathrm{sys, PEPSI}}$ & km s$^{-1}$ & $-17.86\pm0.044$ & This work \\
    $v_{\mathrm{sys, HARPS-N}}$ & km s$^{-1}$ & $-17.15\pm0.11$ & This work \\
    & & $-17.74 \pm 0.11$ & \citealt{Hoeijmakers2019} \\
    & & $-19.819 \pm 0.024$ & \citealt{Borsa2019} \\
    $v_{sys, \mathrm{TRES}}$ & km s$^{-1}$ & $-18.97\pm0.12$ & This work \\
    & & $-20.567 \pm 0.1 $ & \citealt{Gaudi2017} \\
    $K_\star$ & km s$^{-1}$ & $0.23\pm0.060$ & This work \\
    & & $0.276 \pm 0.079$ & \citealt{Gaudi2017} \\
    & & $0.293 \pm 0.032 $ & \citealt{Borsa2019} \\
    $K_{\mathrm{p}}$ & km s$^{-1}$& $239.07^{+5.83}_{-5.79}$ & This work \\
    & & $268.7^{+6.2}_{-6.4}$ & \citealt{Yan2018} \\
    & & $234.24 \pm 0.90 $ & \citealt{Hoeijmakers2019} \\
    & & $241.5^{+3}_{-2} $ & \citealt{Pino2020} \\
    $M_\star$ & $\mathrm{M_\odot}$ & $2.11\pm0.78$ & This work \\
    & & $2.52^{+0.25}_{-0.20}$ & \citealt{Gaudi2017} \\
    & & $3.00 \pm 0.21$ & \citealt{Yan2018} \\
    & & $ 1.978 \pm 0.023$ & \citealt{Hoeijmakers2019} \\
    $m_{\mathrm{p}}$ & $\mathrm{M_J}$ & $2.17 \pm 0.56$ & This work \\
    & & $2.88 \pm 0.84$ & \citealt{Gaudi2017} \\
    & & $3.23 \pm 0.94$ & \citealt{Yan2018} \\
    & & $2.44 \pm 0.70$ & \citealt{Hoeijmakers2019} \\
    $\rho_\star$ & g $\mathrm{cm^{-3}}$ & $0.29\pm0.17$ & This work \\
    & & $0.2702 \pm 0.0029$ & \citealt{Gaudi2017} \\
    \hline
    \end{tabular}
\label{tab:revised_params}
\end{table}

\par Since KELT-9 is a rapidly rotating A0 star ($v\sin{i} = 111.4$ km s$^{-1}$), we recover our own measurements of $K_\star$ and $v_{\mathrm{sys}, i}$ by performing LSD on our time-resolved stellar spectra to recover the rotational broadening kernel (with in-transit observations affected by the Rossiter-McLaughlin effect) of the star at each time of observation. The LSD procedure allows for tunable regularization of the recovered line profile, a feature which cross-correlation does not accomplish. The rotational kernel is centered on the star's radial velocity at the time of observation, which we can fit for using an analytical kernel defined according to the star's $v\sin{i}$. These velocities over time can be fit with the orbital RV equation to determine $K_\star$ and $v_{\mathrm{sys}, i}$. This procedure was previously adopted in \citealt{Borsa2019}.

\par For the first step, we refer to the LSD procedure provided in \citealt{Kochukhov2010} with the modification for regularization given in \citealt{Wang2017}. In our application of this technique, $\mathbf{Y}^0$ is a logarithmically-sampled, $n$-element vector of the observed rotationally broadened stellar spectrum and $\mathbf{F}$ is a template stellar spectrum of the corresponding $T_{\mathrm{eff}}$. $\mathbf{F}$ has the same logarithmic wavelength sampling as the observed spectrum but is not rotationally broadened. We generate $\mathbf{F}$ by inputting the VALD3 linelist for a $T_{\mathrm{eff}} = 10170$ K star into the IDL software \texttt{Spectroscopy Made Easy (SME)} at 21 different limb-darkening angles \citep{Valenti1996, Valenti2012}. We treat the stellar disk as a pixelated grid of $0.01 R_\star \times 0.01 R_\star$ cells to: 1) interpolate in limb-darkening angle to generate the corresponding spectrum for each cell, 2) add up the spectra of each cell, and 3) continuum-normalize to generate the disk-integrated stellar spectrum $\mathbf{F}$. 
\par The deconvolution process recovers the rotational broadening kernel that transforms $\mathbf{F}$ into $\mathbf{Y}^0$. It can be described through matrix multiplication of 1) the cross-correlation between a line mask, $\mathbf{M}$, and the observed spectrum and 2) the inverse of the autocorrelation of the line mask modified by regularization. This is mathematically represented by the following equation (analogous to equation 1 in \citealt{Wang2017} assuming homoscedasticity):
\begin{equation}
    \label{eq:deconvolution}
    \mathbf{Z}(v_i) = (\mathbf{M}^{\mathrm{T}} \cdot \mathbf{M} + \Lambda \mathbf{R})^{-1} \cdot \mathbf{M}^{\mathrm{T}} \cdot \mathbf{Y}^0
\end{equation}
where $\mathbf{Z}(v_i)$ is the m-element vector of the deconvolved line profile (our output of interest). $\mathbf{M}$ is an $n \times m$ line mask constructed by expanding the template spectrum into its corresponding Toeplitz matrix; we adopt the definition provided in \citealt{Donati1997}. For regularization, $\Lambda$ is a regularization parameter, and $\mathbf{R}$ is the $m \times m$ matrix of first-order Tikhonov regularization; we adopt Equation 16 in \citealt{Donatelli2014} for the form of $\mathbf{R}$. The velocities $v_i$ corresponding to the elements in the line profile $\mathbf{Z}(v_i)$ are determined by the wavelength shift of each row in the $\mathbf{M}$ matrix relative to the template spectrum and should be linearly-spaced since we constructed a template spectrum that is logarithmically-spaced in wavelength. 

\par We apply this deconvolution algorithm to all of our out-of-transit observations since the in-transit kernels contain a deficiency in the rotationally-broadened line profile due to the Rossiter-McLaughlin effect. As the planet transits, it blocks a region of the stellar disk and the contribution from this portion of the stellar disk is not added to the integrated stellar disk spectrum. This manifests as a deficiency in the rotational broadening kernel at the radial velocity of that portion of the stellar disk based on stellar rotation. We pooled together our PEPSI data with publicly available HARPS-N data and TRES data from the discovery paper \citep{Gaudi2017} to increase our sample size and out-of-transit coverage (see Table \ref{tab:datasets}).

\begin{figure*}[t]
    \begin{minipage}{0.40\linewidth}
        \centering
        \includegraphics[width=\textwidth]{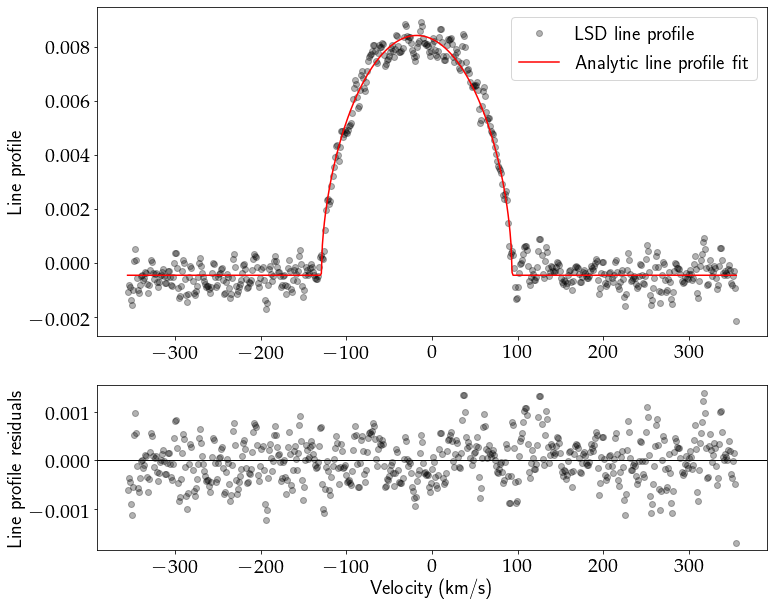} \\
        (a)
    \end{minipage}\hfill
    \begin{minipage}{0.59\linewidth}
        \centering
        \includegraphics[width=\textwidth]{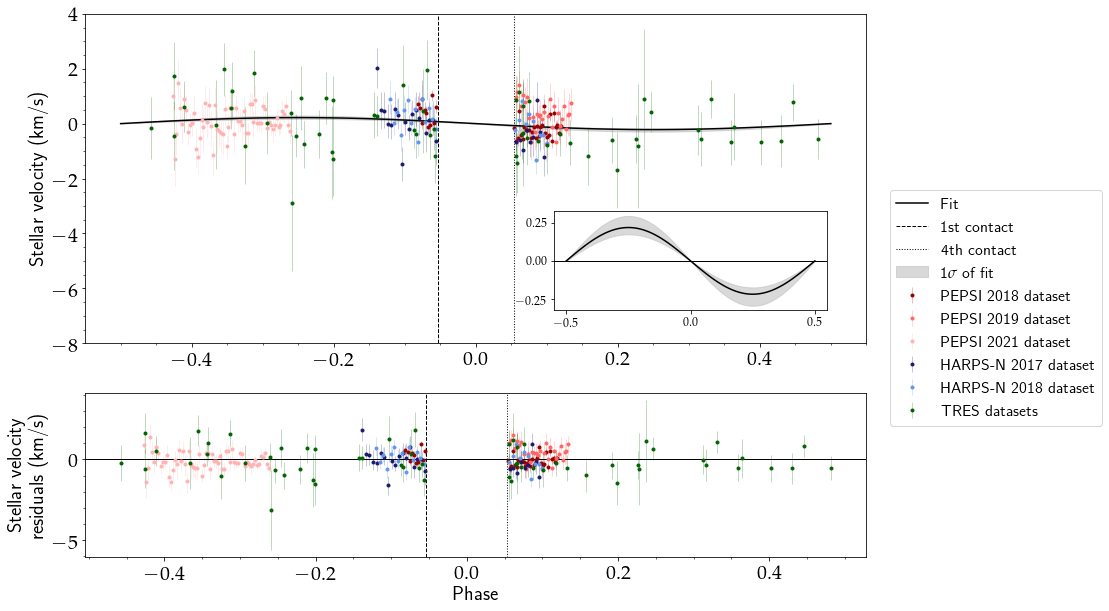} \\
        (b)
    \end{minipage}
    \caption{(a) Example of least-squares deconvolution of a stellar line profile with an analytic rotational kernel fit to centroid stellar radial velocity. (b) Stellar radial velocity curve of the KELT-9 system derived from fitting a circular orbital solution to the out-of-transit stellar radial velocities. The data points are shifted by the best-fit systemic velocities of their corresponding datasets. The inset in the top panel zooms in on the range of velocities along the y-axis spanned by the best-fit orbital solution.}
    \label{fig:stellarRV}
\end{figure*}



\subsubsection{Analytic rotational kernel fit to stellar line profiles}
Upon generating the observational line profiles, we fit each one with an analytic rotational kernel model to determine their centers and recover an orbital RV curve of the star. We use the analytical expression for a rotational broadening kernel given in \citealt{gray_2005} (Equation 18.14), which is adapted for our application as follows:
\begin{equation}
    \label{eq:rot_kernel_analytic}
    \mathbf{G}(v_i, t) = c_1 \Big[1 - \Big(\frac{v_i - v_\star(t)}{v\sin{i}} \Big)^2 \Big]^{1/2} + c_2 \Big[1 - \Big(\frac{v_i - v_\star(t)}{v\sin{i}} \Big)^2 \Big]
\end{equation}
where $\mathbf{G}(v_i)$ is the analytical rotational kernel defined over a range of velocities $v_i$; $v_\star(t)$ is the centroid of the rotational kernel and represents the radial velocity of the star at the time $t$ of a given observation; $v\sin{i}$ is the width of the kernel and represents the sky-projected rotational velocity of the star (we used the value from \citealt{Gaudi2017} of 111.4 km s$^{-1}$ to ensure the same value was used across observations; the residuals in the left panel of  Figure \ref{fig:stellarRV} suggest this value sufficiently matches the data); and $c_1$ and $c_2$ are constants defined in terms of $v\sin{i}$ and the linear limb-darkening coefficient of the star, $\epsilon$, as
\begin{equation}
c_1 = \frac{2(1-\epsilon)}{\pi v\sin{i} (1-\epsilon/3)}
\end{equation}
and
\begin{equation}
c_2 = \frac{\epsilon}{2 v\sin{i} (1-\epsilon/3)}
\end{equation}

\par For each observation, we apply least-squares fitting from \texttt{scipy} \footnote{\texttt{https://www.scipy.org/index.html}} \citep{scipy2020} to determine the best-fit values of $v_\star$, $\epsilon$, a multiplicative scaling factor (rescales the analytical kernel to match the scaling of the empirical deconvolved line profiles), and an additive offset (the empirical deconvolved line profiles may not have a baseline centered at 0 due to a lack of flux conservation between the template and observed spectra). Of these parameters, we are most interested in $v_\star$, which provides measurements of the star's radial velocity over time. We estimated radial velocity errors by bootstrapping the residuals of the flat region of the deconvolved kernel ($|$velocity$|\ > v \sin{i}$ in the left panel of Figure \ref{fig:stellarRV}), adding the samples to the best-fit model kernel, and refitting the line profile. We repeated this bootstrapping procedure to obtain 1000 resampled values of $v_\star$ for each observation and took the \nth{16} percentile and \nth{84} percentile samples as the $\pm1\sigma$ errors on the best-fit $v_\star$. 

\par We also tested fixing the limb-darkening parameter to $\epsilon=0.3356$ based on the stellar parameters that best describe KELT-9 in the limb-darkening tables provided by \citealt{Claret2017}. Upon refitting the rotational broadening profiles, we find that the differences in the recovered stellar RVs when $\epsilon$ is fixed are not significant enough to affect the resulting stellar orbital parameters (discussed in Section \ref{sec:stellarOrbitalSolution}) within errors.

\subsubsection{Stellar orbital solution}
\label{sec:stellarOrbitalSolution}
\par To obtain the stellar orbital parameters, we fit a circular orbital RV solution to our time-resolved measurements of stellar velocity, which has the following form:
\begin{equation}
    v_\star(t) = K_\star \sin \Big( \frac{t - t_0}{p} \Big) + v_{\mathrm{sys}, i}
\end{equation}
where $v_\star(t)$ is the stellar radial velocity as a function of time, $K_\star$ is the stellar RV semi-amplitude (stellar orbital velocity for a circular orbit), $t_0$ is the mid-transit time, $P$ is the orbital period, and $v_{\mathrm{sys}, i}$ is the systemic velocity of a given instrument. We apply the Markov Chain Monte Carlo (MCMC) method to sample the parameter space  using the \texttt{emcee}\footnote{\texttt{https://emcee.readthedocs.io/en/stable/}} code \citep{Foreman-Mackey2013}. We fit three common parameters that apply across all datasets: mid-transit time ephemeris ($t_0$), period ($p$), and stellar RV semi-amplitude ($K_{\mathrm{p}}$). We also fit three individual systemic velocities that singly apply to their corresponding datasets: $v_{\mathrm{sys, PEPSI}}$, $v_{\mathrm{sys, HARPS-N}}$, and $v_{\mathrm{sys, TRES}}$. Assuming a fixed $v_{\mathrm{sys}, i}$ across all datasets for a given instrument is valid since all three instruments are wavelength-calibrated against a ThAr reference which guarantees instrumental RV stability. 

\begin{table*}[t]
\centering
\caption{Stellar orbital solution MCMC priors.}
    \begin{tabular}{lll}
    \hline
    \hline
    Parameter & Units & Prior\\ 
    \hline
    \hspace{3mm} $t_0$ & $\mathrm{BJD_{TDB}}$ & Gaussian prior \\
    & & Center: 2458566.436560 \\
    & & Amplitude: 0.000048 \\
    \hspace{3mm} $p$ & days & Gaussian prior \\
    & & Center: 1.48111890 \\
    & & Amplitude: 0.00000016 \\
    \hspace{3mm} $K_\star$ & km $\mathrm{s}^{-1}$ & Linearly uniform prior \\
    & & Lower bound: -1 \\
    & & Upper bound: 0\\
    \hspace{3mm} $v_{\mathrm{sys, PEPSI}}$ & km $\mathrm{s}^{-1}$ & Linearly uniform prior \\
    & & Lower bound: -100 \\
    & & Upper bound: 0\\
    \hspace{3mm} $v_{\mathrm{sys, HARPS-N}}$ & km $\mathrm{s}^{-1}$ & Linearly uniform prior \\
    & & Lower bound: -100 \\
    & & Upper bound: 0\\
    \hspace{3mm} $v_{\mathrm{sys, TRES}}$ & km $\mathrm{s}^{-1}$ & Linearly uniform prior \\
    & & Lower bound: -50 \\
    & & Upper bound: 50\\
    \hline
    \end{tabular}
\label{tab:star_MCMC_priors}
\end{table*}

\par We apply linearly uniform priors on all model parameters except $t_0$ and $p$ for which we use Gaussian priors centered on the mid-transit time ephemeris and period respectively as provided in Table \ref{tab:system_params} ($T_0$ and $P$) with their corresponding errors as the standard deviations of the Gaussians. See Table \ref{tab:star_MCMC_priors} for priors. The priors $T_0$ and $P$ are updated ephemerides for the KELT-9 b system refitted with \texttt{EXOFASTv2} \citep{Eastman2019} to include recent TESS observations \citep{TESS2015} in addition to the follow up lightcurves and TRES RVs from the discovery paper. The updated ephemerides are critical for this analysis since the observations span multiple years; the error propagation using the original ephemerides can be as large as 4.8 minutes for the PEPSI 2019 dataset. Introducing the TESS 2019 data lightcurves to the global fit improves the precision of the ephemerides and eliminates the issue of "stale" ephemerides by spanning a broad temporal baseline.

\par We evaluate the goodness-of-fit for a given model using the log-likelihood $\ln{\cal L} \equiv -\frac{\chi^2}{2}$. Here $\chi^2$ is the statistical chi-squared, defined as $\chi^2 = \sum \Big(\frac{\mathrm{data}-\mathrm{model}}{\mathrm{data\ errors}}\Big)^2$, of the model. To ensure the log-prior is on a comparable scale as the log-likelihood, we scale the priors by the number of elements in the observed stellar RV curve before summing the log-prior and log-likelihood in the log-posterior. We run the \texttt{emcee} sampler with 10 walkers until the chain length is at least 100 times the estimated autocorrelation time and the estimated autocorrelation time has changed by less than 1\%, checking every 500 steps. Under these criteria, the posterior distributions of all parameters appear sufficiently Gaussian or converged. See the right panel of Figure \ref{fig:stellarRV} for the observed and best-fit model stellar RV curve. Table \ref{tab:revised_params} reports the fitted systemic velocities and stellar RV semi-amplitude from this analysis.

\subsection{Planet orbital properties}
To obtain the orbital properties of the planet, we generate transmission spectra which feature the planet's atmospheric absorption track during transit as well as secondary effects due to the geometry of the planet's transit across a non-uniform stellar disk. We simultaneously model both of these signatures and apply a Bayesian framework for fitting the data with MCMC; the most relevant outputs of this procedure is the orbital velocity of the planet, $K_{\mathrm{p}}$, which we will use later in this work to constrain the dynamical mass of the planet.

\subsubsection{Transmission spectrum construction}
\label{sec:transSpec}
\par
\begin{figure}[t]
    \hspace{-3mm}
    \includegraphics[width=0.5\textwidth]{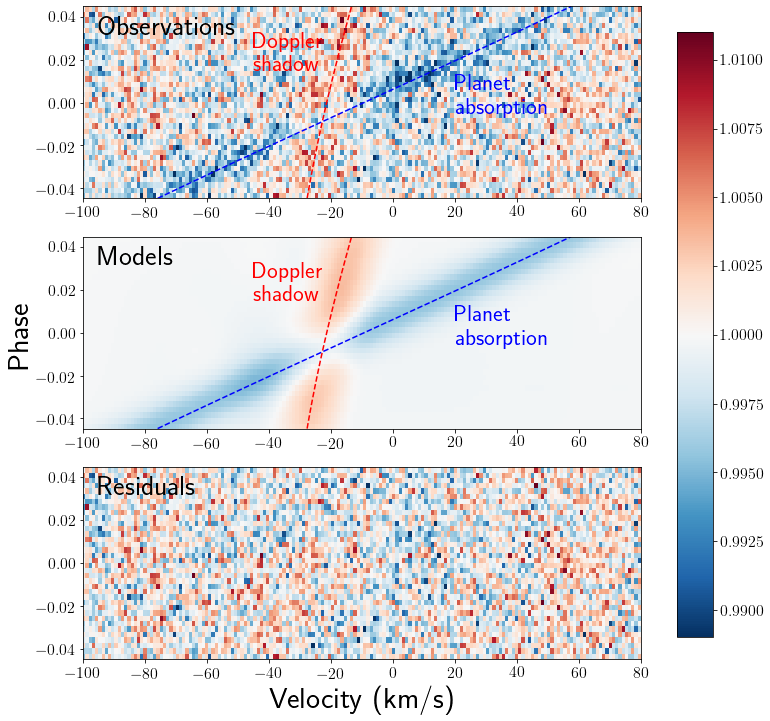}\\
    \caption{2D map of transmission spectra over the course of KELT-9 b's transit for the PEPSI 2018 dataset; the blue track is formed by the planet's atmospheric absorption while the red track is the Doppler shadow from the Rossiter-McLaughlin effect.Top panel displays fully in-transit observations. Middle panel shows the best-fit model from MCMC sampling, with the Doppler shadow and CLV determined from numerical modelling of the planet's transit using SME stellar models, while the planet absorption track is a uniform Gaussian signal shifted in velocity according to the best-fit orbital motion of the planet, systemic velocity, and best-fit day-to-nightside winds. The bottom panel shows the residuals (data - model).}
    \label{fig:transSpec}
\end{figure}

 We constrain the RV semi-amplitude of the planet, $K_{\mathrm{p}}$, and global day-to-nightside winds, $v_{\mathrm{wind}}$, from a line-by-line analysis of the atmospheric absorption signature seen in our transmission spectra of KELT-9 b from PEPSI. To extract the planet's atmospheric absorption signature, we chose to focus on 6 Fe II lines in the wavelength range between 4915--5400 {\AA} to avoid telluric contamination and broad features such as the H$\beta$ line around 4861.4 {\AA}. The outputs from the PEPSI pipeline are continuum-normalized stellar spectra, which we interpolate onto the same logarithmically-spaced wavelength grid constructed such that absorption features are as well-sampled as they are by the original wavelength grid. This enables uniform-spacing in velocity across all observations. We constructed an empirical combined stellar spectrum by taking the average of the out-of-transit spectra, which were identified using updated ephemerides of the KELT-9 b system (see Table \ref{tab:system_params}). To remove the stellar component, we divide all spectra by the combined stellar spectrum. The residual transmission spectra contain the planet's signature as well as secondary geometric effects caused by the Rossiter-McLaughlin Effect (RME) \citep{Rossiter1924, McLaughlin1924, Queloz2000, Ohta2005, Gaudi2007} and center-to-limb variation (CLV) \citep{Stenflo2015, Yan2015, Yan2018}, which are both consequences of non-uniformity across the stellar disk (see the Doppler Shadow in Figure \ref{fig:transSpec}, a byproduct of RME). These features are distinguishable because they span different regions of velocity space and contribute opposing signs to the flux map.

\par Upon inspection of our transmission spectra in the wavelength range between 4915--5400 \text{\AA} for planet atmospheric signatures, we noticed that Fe II lines presented the strongest atomic absorption signatures (apart from Balmer lines, which are significantly broader as well). Previous studies of KELT-9 b's atmosphere with transmission spectroscopy revealed a diverse array of Balmer lines ranging from H$\alpha$ to H$\zeta$ \citep{Yan2018, Cauley2019, Wyttenbach2020}. Due to the extreme heating of KELT-9 b's atmosphere, traces of heavy metals, both neutral and ionized, have also been found either by directly investigating the atomic absorption lines \citep{Cauley2019} or by harnessing the cross-correlation technique to boost the absorption signal \citep{Hoeijmakers2019}. Like \citealt{Cauley2019}, we find that the strongest metal features in our selected wavelength range are produced by Fe II; although signatures of Fe I, Ti II, and Mg I can be observed without cross-correlation in our PEPSI datasets, we have chosen to focus on the Fe II lines to ensure that our measurement of $K_{\mathrm{p}}$ is reliably derived from narrow absorption features that have the highest signal-to-noise.

\par For each of our six selected Fe II lines, we focus on a tightly restricted span of wavelengths that completely encapsulates the planet absorption and Doppler shadow signatures for the observations that are fully in-transit (between 2nd and 3rd contact). Thus we have a flux map of the fully in-transit observations for each Fe II line. Wavelength is on the x-axis and orbital phase is on the y-axis. The values in the map correspond to the continuum-normalized transmission spectrum fluxes during transit. 

\par We then generate a template spectrum of the Fe II species the planet's atmosphere using \texttt{petitRADTRANS}\footnote{\texttt{https://petitradtrans.readthedocs.io/en/latest/}} \citep{Molliere2019}, a radiative transfer code for modelling transmission and emission spectra of planetary atmospheres. We note that at the time of writing, the Fe II opacities used by \texttt{petitRADTRANS} are in air, unlike most other species which are given in vacuum; hence we did not need to perform any wavelength correction to match the template spectrum wavelengths with our observations. The planetary parameters we adopt for $R_{\mathrm{p}}$, $R_\star$, $m_{\mathrm{p}}$, $\log{g_{\mathrm{p}}}$, and $T_{\mathrm{eq}}$ are provided in Table \ref{tab:system_params}. Additional parameters that are necessary for generating a transmission spectrum in \texttt{petitRADTRANS} are available in Table \ref{tab:petit_params}. Abundances were chosen to be close to solar \citep{Palme2014}, while the pressure structure was constructed to encompass the region of the atmosphere that most strongly affects line formation in transmission spectra. It is not essential to accurately model the specific shape of lines in the resulting template spectrum of the planet's atmosphere, as the line centers are the most relevant quantity for our analysis of the planet's orbital velocity through the Doppler effect. 

\par
\begin{figure*}[t]
    \includegraphics[width=1\textwidth]{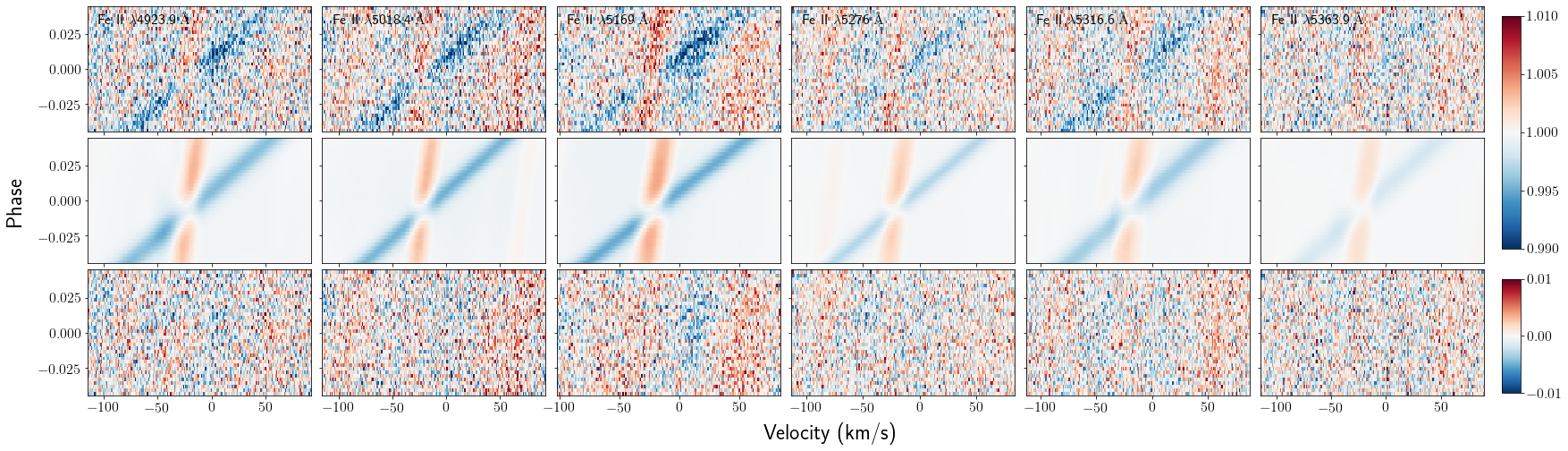}
    \caption{Expanded version of Figure \ref{fig:transSpec} displaying the six Fe II absorption lines chosen for fitting $K_{\mathrm{p}}$ (PEPSI 2018 dataset).}
    \label{fig:FeIIlines}
\end{figure*}

 With the wavelengths of the Fe II line centers from the template spectrum, we can convert our flux maps from wavelength space to velocity space with the simplified Doppler effect for non-relativistic speeds:
\begin{equation}
    v \approx c\frac{\lambda-\lambda_0}{\lambda_0}
    \label{eq:dopplerVelocity}
\end{equation}
where $\lambda_0$ is the line center corresponding to the Fe II line that generates the absorption feature in a given flux map. We shift our flux maps to the rest-frame of the star-planet system by subtracting the systemic velocity calculated in Section \ref{sec:stellarOrbitalSolution} from the velocity grid of each flux map. The resulting flux maps of each Fe II line is presented in Figure \ref{fig:FeIIlines}. 

\begin{table}[b]
\centering
\caption{Additional \texttt{petitRADTRANS} parameter inputs}
    \begin{tabular}{lll}
    \hline
    \hline
    Parameter & Units&  Value  \\ 
    \hline
    Equilibrium temperature & K & 4050 \\
    Internal temperature & K & 100 \\
    Pressure range & bar & $10^{-10}$ -- $10^2$ \\
    Reference pressure & bar & $10^{-9}$ \\
    Infrared atmospheric opacity & & 0.01 \\
    Ratio between optical and IR opacity & & 0.4 \\
    Abundances: \\
    \hspace{3mm} Fe II  & & 0.00125 \\ 
    \hspace{3mm} $\mathrm{H_2}$ & & 0.748 \\
    \hspace{3mm} He & & 0.250 \\
    \hline
    \end{tabular}
\label{tab:petit_params}
\end{table}

\subsubsection{Secondary effect modelling}
\label{sec:secondaryEffects}

\par We follow the numerical approach to modelling RME and CLV presented in \citealt{Casasayas-Barris2019} (originally presented in \citealt{Yan2015} and \citealt{Yan2017}, modified for planet radius interpolation in \citealt{Casasayas-Barris2019}). We obtain linelists corresponding to the properties of the KELT-9 host star from VALD3 \citep{Pakhomov2017}. Known stellar properties from \citealt{Gaudi2017} are input into the IDL software \texttt{Spectroscopy Made Easy (SME)} to model the host star spectrum using the VALD3 \citep{Pakhomov2017} linelists at 21 different limb-darkening angles \citep{Valenti1996, Valenti2012}. We model the planet's transit across a pixelated grid of $0.01 R_\star \times 0.01 R_\star$ cells that make up the stellar disk. The sky-projected position of the planet is computed according to Equations \ref{eq:pos_1_x} and \ref{eq:pos_1_y}, which are analogous to Equations 7, 8, and 10 in \citealt{CollierCameron2010} (with the sign error in their Equation 10 corrected; this error was originally noted in \citealt{Eastman2019}):
\begin{align}
    x &= x_{\mathrm{p}} \cos{\lambda} + y_{\mathrm{p}} \sin{\lambda} 
    \label{eq:pos_1_x} \\
    y &= -x_{\mathrm{p}} \sin{\lambda} + y_{\mathrm{p}} \cos{\lambda}
    \label{eq:pos_1_y}
\end{align}
where 
\begin{align}
    x_{\mathrm{p}} &= a \sin{\frac{2\pi(t - t_0)}{P}} 
    \label{eq:pos_2_x} \\
    y_{\mathrm{p}} &= a \cos{\frac{2\pi(t - t_0)}{P}} \cos{i_{\mathrm{orbit}}}
    \label{eq:pos_2_y}
\end{align}
In Equations \ref{eq:pos_1_x}--\ref{eq:pos_2_y}, $\lambda$ is the sky-projected obliquity, $a$ is the semi-major axis of the planet's orbit, $t_0$ is the mid-transit time, $P$ is the orbital period of the planet, $i_{\mathrm{orbit}}$ is the orbital inclination of the planet, and $t$ is the time of a given observation; the system parameters related to these equations can be found in Table \ref{tab:system_params}. We generate a grid of orbital phases that Nyquist samples the phases of the observations and for each phase in the grid, we identify which cells are unobscured by the planet's transit. We interpolate in limb-darkening angle to generate a spectrum for each unobscured cell and shift the spectrum in velocity according to the cell's radial velocity as determined by its location on the stellar disk and the $v\sin{i}$ of the star. Then we add up the velocity-shifted spectra of each unobscured cell, continuum-normalize, and divide by the out-of-transit model spectrum (the sum of the spectra of all cells interpolated in limb-darkening angle and shifted in radial velocity, then continuum-normalized) to generate the disk-integrated transmission spectrum (excluding the planet's absorption signature) for each phase. In this manner, we produce a 2D map of spectra over different orbital phases that encapsulate both the Rossiter-McLaughlin effect and the center-to-limb variation that perturb the transmission spectra during the planet's transit. We generate a grid of such models for a range of planet radii spanning $0.7 R_{\mathrm{p}} < R < 2.5 R_{\mathrm{p}}$ as done in \citealt{Casasayas-Barris2019}.

\subsubsection{Line-by-line analysis of Fe II lines}
\label{sec:line-by-line}
\begin{figure}[b]
    \hspace{-5mm}
    \includegraphics[width=0.5\textwidth]{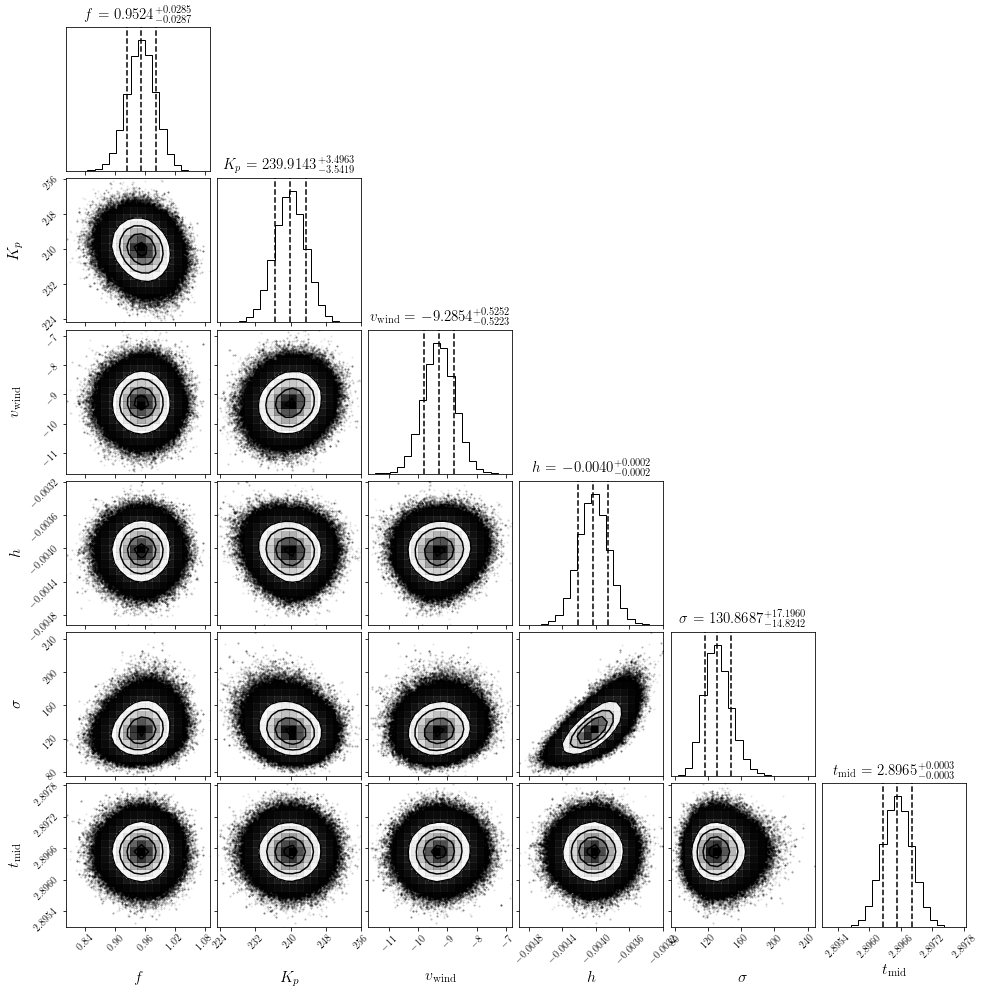}
    \caption{Example of a corner plot from MCMC fitting of the Fe II $\lambda4923.9$ \text{\AA} flux map.}
    \label{fig:MCMC_corner}
\end{figure}

We apply a procedure analogous to the Markov chain Monte Carlo (MCMC) fitting of the observed transmission spectra presented in \citealt{Casasayas-Barris2019} using \texttt{emcee}. To avoid additional asymmetries and velocity offsets from equatorial jets and rotation, we exclude observations during ingress and egress and instead focus on fully in-transit observations such that any velocity shifts from atmospheric dynamics can be treated as a constant blueshifted offset from day-to-nightside winds. We fit for the following free parameters: the effective planet radius factor in RME/CLV modelling ($f$), the RV semi-amplitude of the planet ($K_{\mathrm{p}}$), the radial velocity of the terminator-averaged atmosphere in the planet's restframe ($v_{\mathrm{wind}}$), the contrast of the planet's Gaussian atmospheric absorption profile ($h$), the standard deviation of the planet's Gaussian atmospheric absorption profile ($\sigma$), and the mid-transit time ($t_0$). See Figure \ref{fig:MCMC_corner} for a sample corner plot.

\par We generate model flux maps according to the input parameters as follows. The parameter $f$ corresponds to a model map of the RM and CLV effects from the grid of models generated in Section \ref{sec:secondaryEffects} by interpolating between models in planet radius according to the value of $f$, then interpolating the resulting map in wavelength (then converted to velocity according to the reference Fe II line of the data map being fitted and applying Equation \ref{eq:dopplerVelocity}) and orbital phase to match the observations, i.e. the single line data maps generated in Section \ref{sec:transSpec}. 

\par Recall that $t_0$ is a free parameter in this fitting procedure (this is not done in \citealt{Casasayas-Barris2019} or \citealt{Yan2017}; we have added this extra free parameter so that we can include the uncertainty of mid-transit time, which is strongly correlated with the parameter $v_{\mathrm{wind}}$, in our analysis. We inject a Gaussian signal in our model maps that varies in velocity with time but remains constant in amplitude and width to represent the planet's absorption signature. The parameters $K_{\mathrm{p}}$, $v_{\mathrm{wind}}$, and $t_0$ determine the center of the planet's atmospheric absorption signal as a function of orbital phase (which is, once again, dependent on the free parameter $t_0$) in our models as follows:
\begin{equation}
    v_{\mathrm{p}}(\phi) = K_{\mathrm{p}} \sin(2\pi\phi) + v_{\mathrm{wind}}
    \label{eq:vwind}
\end{equation}
The strength and width of the planet's atmospheric absorption signal is dependent on the free parameters $h$ and $\sigma$, resulting in the following form for the planet's absorption signature as a function of the radial velocities $v$ and orbital phases $\phi$ that correspond with the observed flux maps:
\begin{equation}
    T_v(\phi) = 1 + h\mathrm{e}^{\frac{v-v_{\mathrm{p}}(\phi)}{2 \sigma}}
\end{equation}
This is analogous to Equation 1 in \citealt{Yan2018}.
\par We adopt a Bayesian framework for sampling the parameter space with MCMC. We apply linearly uniform priors on all model parameters except $t_0$, for which we use a Gaussian prior centered on the mid-transit time value provided in Table \ref{tab:system_params} ($T_0$, propagated to the epoch of the corresponding dataset being fitted) with a standard deviation that matches the propagated mid-transit time error. See Table \ref{tab:planet_MCMC_priors} for priors. As before, we use $-\frac{\chi^2}{2}$ as the log-likelihood of a given model and scale the priors by the number of elements in the observed flux map (which should be the same as the number of elements in the model flux map) before summing the log-prior and log-likelihood in the log-posterior. We run the \texttt{emcee} sampler with 10 walkers until the the estimated autocorrelation time is 1.5\% of the chain length and the estimated autocorrelation time has changed by less than 1\% , checking every 500 steps. Under these criteria, the posterior distributions of all parameters appear sufficiently Gaussian or converged.

\begin{table*}[t]
\centering
\caption{Planet orbital solution MCMC priors (same for both line-by-line and cross-correlation analysis).}
    \begin{tabular}{llp{13cm}}
    \hline
    \hline
    Parameter & Units &  Prior \\ 
    \hline
    \hspace{3mm} $f$ & & Uniform prior \\
    & & Lower bound: 0.7 \\
    & & Upper bound: 2.5 \\
    \hspace{3mm} $K_{\mathrm{p}}$ & km $\mathrm{s}^{-1}$ & Uniform prior \\
    & & Lower bound: 100 \\
    & & Upper bound: 350 \\
    \hspace{3mm} $v_{\mathrm{wind}}$ & km $\mathrm{s}^{-1}$ & Uniform prior\\
    & & Lower bound: -50 \\
    & & Upper bound: 50 \\
    \hspace{3mm} $h$ & & Uniform prior\\
    & & Lower bound: -1 \\
    & & Upper bound: 0 \\
    \hspace{3mm} $\sigma$ & velocity pixels & Uniform prior \\
    & & Lower bound: 0 \\
    & & Upper bound: 1000 \\
    \hspace{3mm} $t_{0}$ & $\mathrm{BJD_{TDB}}$ & Gaussian prior \\
    & & Center: 2458566.436560 propagated to epoch of dataset \\
    & & Width: 0.000048  propagated (according to error on orbital period) to epoch of dataset \\
    \hline
    \end{tabular}
\label{tab:planet_MCMC_priors}
\end{table*}

\par The main goal of the model-fitting for the purposes of this work is to recover a value for $K_{\mathrm{p}}$; $v_{\mathrm{wind}}$, or day-to-nightside winds, will be the topic of a future work. The last 25\% of samples are extracted from the 10 walkers for a given Fe II line and aggregated into one chain for each Fe II line. We weight each sample from all six chains according to the signal-to-noise ratio (S/N) of the Fe II line flux map it corresponds to. Signal of a given flux map is the 50th percentile sample of $h$ in the flux map's chain. Noise is the standard deviation of the residuals, i.e. best-fit model subtracted from the flux map, where the best-fit model is a model generated by the 50th percentile parameters from the chain of the flux map. We combine the weighted chains to determine a representative value for the quantity of interest. We recover a representative value of $K_{\mathrm{p}}$ by taking the S/N-weighted 50th percentile sample across all chains. The lower and upper 1$\sigma$ values are the S/N-weighted 16th percentile and 84th percentile samples respectively. See Figure \ref{fig:Kp} for the best-fit $K_{\mathrm{p}}$ for each line analyzed in both the PEPSI datasets as well as the representative value (dashed black line) from combining the MCMC analysis of all the lines. Literature values are provided for comparison.

\subsubsection{Planet orbital properties: Cross-correlation analysis of Fe II lines}
\label{sec:cross_corr}

\begin{figure*}[t]
    \begin{minipage}{0.5\linewidth}
        \centering
        \includegraphics[width=\textwidth]{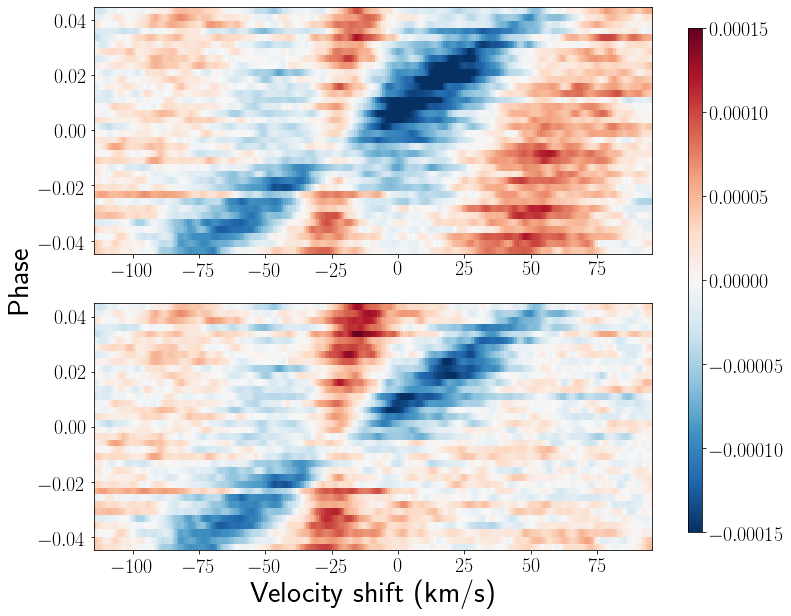} \\
        (a)
    \end{minipage}\hfill
    \begin{minipage}{0.5\linewidth}
        \centering
        \includegraphics[width=\textwidth]{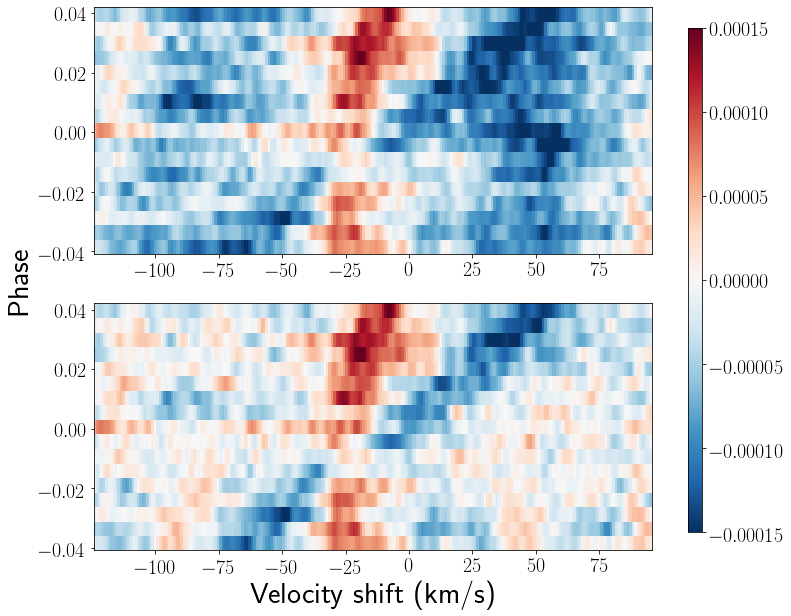} \\
        (b)
    \end{minipage}
    \caption{Cross-correlated flux maps before artifact correction in the top panel and after artifact correction in the bottom panel for the (a) PEPSI 2018 dataset and (b) HARPS-N 2017 dataset.}
    \label{fig:pulsations}
\end{figure*}

\begin{figure*}[t]
    \includegraphics[width=1\textwidth]{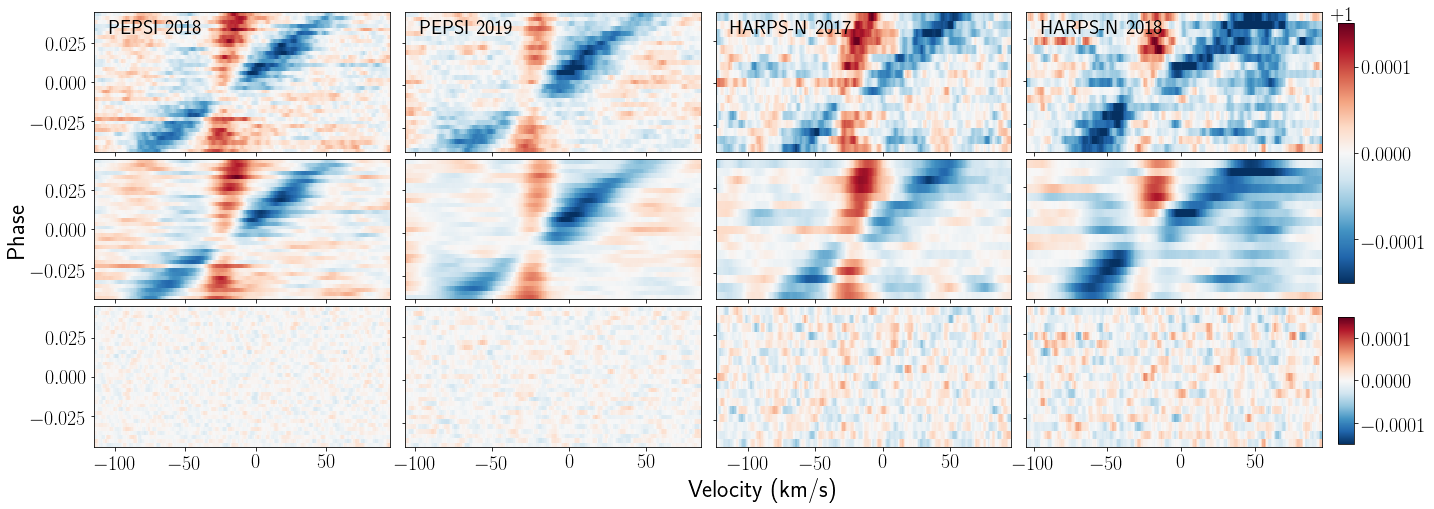}
    \caption{Same as Figure \ref{fig:FeIIlines} except displaying CCF maps for each dataset.}
    \label{fig:FeIICCFs}
\end{figure*}

Since the SNR for the HARPS-N data is significantly lower than the PEPSI data, individual lines are noisier in the HARPS-N observations. Cross-correlation is a technique that can boost the SNR by adding up contributions from matching signals between an observed and template spectrum across multiple features. For comparison with the publicly available archival HARPS-N observations of KELT-9 b, we cross-correlate our transmission spectra with a template Fe II template spectrum (we will call it $f(\lambda)$ and the observed flux map $F(\lambda, \phi)$). We cross-correlate our PEPSI and HARPS-N transmission spectra between 4950 {\AA} to 5400 {\AA} (avoiding the broad H$\mathrm{\beta}$ feature) with the Fe II template generated in \texttt{petitRADTRANS} from \ref{sec:transSpec} over a sufficient range of velocities to span the atmospheric absorption signature. The cross-correlation function we adopt is defined as follows:
\begin{equation}
    CCF(v, \phi) = \sum_{\lambda_i=\lambda_{\mathrm{min}}}^{\lambda_{\mathrm{max}}} f(\lambda_i, v)F(\lambda_i, \phi)
\end{equation}
where $f(\lambda, v)$ is the template spectrum shifted by velocity $v$. Likewise, we cross-correlate our grid of RME/CLV models from \ref{sec:secondaryEffects} with the same Fe II template. Consequently, we obtain CCF maps analogous to Figure \ref{fig:transSpec} that can be fit with the same MCMC procedure as given in \ref{sec:line-by-line}. See Figure \ref{fig:FeIICCFs} for observed and model CCF maps and the corresponding residuals for all 4 transmission spectroscopy datasets.

\par Our cross-correlated flux maps display distinct features that are not attributable to the planet's atmospheric absorption track or the RME/CLV perturbations. In particular, our cross-correlation procedure reproduces the artifacts described in \citealt{Hoeijmakers2019} as "stellar pulsations" in the HARPS-N datasets (see Figure \ref{fig:pulsations}). We model these artifacts as individual Gaussian features which have time-variable amplitudes, standard deviations, and centroids that gradually change across observations. We fit these features using non-linear least-squares optimization and subtract the model from the observed cross-correlation flux map. Since these artifacts overlap in velocity-phase space with the atmospheric absorption track, we first subtract a preliminary fit of the planet's absorption and the Doppler shadow before removing the artifacts. 

\par We also adopt a correlated noise model in our MCMC fitting procedure of the cross-correlated maps using the Gaussian process regression library \texttt{george}\footnote{\texttt{https://george.readthedocs.io/en/latest/}} (see panels in the second row of Figure \ref{fig:FeIICCFs}). Without this added component to the model, the parameter errors are notably underestimated. We assume a 2-dimensional Mat\'ern-3/2 covariance kernel, resulting in one amplitude parameter and two length parameters in addition to our original model parameters to fit with MCMC. This additional step does not yield parameter posterior distributions that are statistically significantly different for the line-by-line analysis. For our cross-correlated data, however, including the Gaussian process to model the covariance structure of the data accurately captures the uncertainty from correlated noise. 

\subsection{Measuring planet and stellar masses}
\label{sec:measuring_mass}

As in \citealt{Snellen2010}, we treat the system as a double-lined spectroscopic binary to determine the planet and stellar masses. In previous sections, we demonstrated that: 1) the absorption lines of the planet's atmosphere during transit capture the planet's orbital velocity while 2) the out-of-transit observations can be deconvolved with a template stellar spectrum to retrieve the stellar orbital velocity. From conservation of momentum, we know that
\begin{equation}
    \label{eq:momentum}
    m_{\mathrm{p}} K_{\mathrm{p}} = M_\star K_\star
\end{equation}
where $m_{\mathrm{p}}$ and $M_\star$ are the masses and $K_{\mathrm{p}}$ and $K_\star$ are the RV semi-amplitudes (or orbital velocity in the case of a circular orbit) of the planet and star respectively. Kepler's 3rd law of orbital motion states
\begin{equation}
    \label{eq:Kepler3}
    P^2 = \frac{4 \pi^2}{G(M_\star + m_{\mathrm{p}})}(a_\star + a_{\mathrm{p}})^3 
\end{equation}
Assuming KELT-9 b is on a circular orbit based on an estimate of its circularization timescale \citealt{Borsa2019}, we can relate the semi-major axes with observables we have previously measured, namely the RV semi-amplitudes of the planet ($K_{\mathrm{p}}$) and star ($K_\star$), orbital inclination ($i$), and the orbital period ($P$):
\begin{align}
    \label{eq:K_p}
    \frac{K_{\mathrm{p}}}{\sin{i}} &= \frac{2\pi a_{\mathrm{p}}}{P} \\
    \label{eq:K_star}
    \frac{K_\star}{\sin{i}} &= \frac{2\pi a_\star}{P}
\end{align}
where $a_{\mathrm{p}}$ and $a_\star$ are the orbital semi-major axes of the planet and star respectively. We can rewrite $M_\star$ in terms of $m_{\mathrm{p}}$ using Equation \ref{eq:momentum} and apply Equations \ref{eq:K_p} and \ref{eq:K_star} to recast the semi-major axes in terms of RV semi-amplitudes and orbital period. This allows us to write our quantity of interest, $m_{\mathrm{p}}$, in terms of purely empirical (and predominantly spectroscopic) observables. Upon doing so, we derive the following expression for planet mass:
\begin{equation}
    \label{eq:m_p}
    m_{\mathrm{p}} = \frac{P K_\star}{2 \pi G K_{\mathrm{p}}} \Big(\frac{K_{\mathrm{p}}}{\sin{i}}\Big)^3 \Big(1 + \frac{K_\star}{K_{\mathrm{p}}}\Big)^2
\end{equation}
Applying this result to conservation of momentum (Equation \ref{eq:momentum}) yields the following expression for stellar mass:
\begin{equation}
    \label{eq:M_star}
    M_\star = \frac{P}{2 \pi G} \Big(\frac{K_{\mathrm{p}}}{\sin{i}}\Big)^3 \Big(1 + \frac{K_\star}{K_{\mathrm{p}}}\Big)^2
\end{equation}
Uncertainties on $m_\mathrm{p}$ and $M_\star$ are derived according to linear propagation of errors (see Appendix, Section \ref{sec:errorProp} for the analytic expressions).

\section{Results}
\label{sec:results}

\subsection{Summary of findings}
We present the revised planetary and stellar parameters resulting from our analysis in Table \ref{tab:revised_params}. Most notable is our measurement of the planetary mass, the first self-consistent, purely empirical measurement of KELT-9 b's dynamical mass, $2.17 \pm 0.56\ \mathrm{M_J}$, by treating the system as a double-lined spectroscopic binary. Another byproduct of this analysis is the mass of the star, which we measure to be $2.11\pm0.78\ \mathrm{M_\odot}$.

\subsection{Comparison with previous literature}
\par We now compare our measurement of the dynamical mass of KELT-9 b against previous literature (see Table \ref{tab:revised_params}). Throughout the following discussion, we define $1\sigma$ by adding in quadrature the errors from two studies of comparison for a parameter. Our measurement of $m_{\mathrm{p}}=2.17 \pm 0.56\ \mathrm{M_J}$ agrees with \citealt{Gaudi2017} and \citealt{Yan2018}, and \citealt{Hoeijmakers2019}. 

\par
\begin{figure}[t]
    \hspace{-5mm}
    \includegraphics[width=0.5\textwidth]{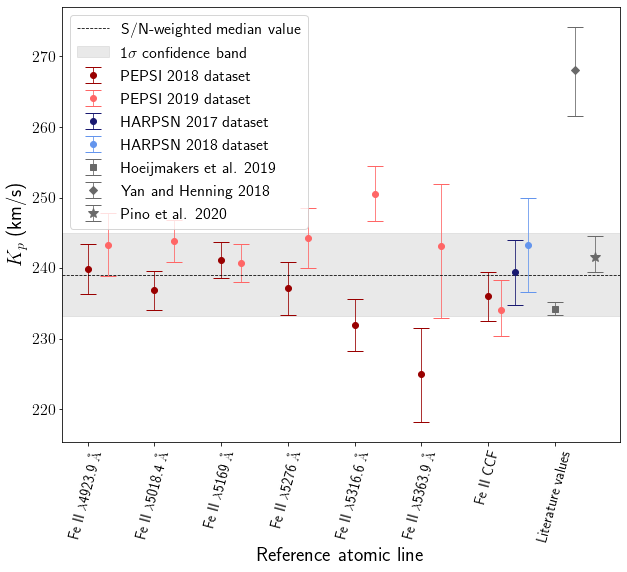}
    \caption{$K_{\mathrm{p}}$ measurements from Fe II line-by-line and cross-correlation analysis with MCMC sampling errors.}
    \label{fig:Kp}
\end{figure}

 The discrepancy in mass measurements across the literature can be largely attributed to the use of $K_\star$ given in \citealt{Gaudi2017} and independent measurements of $K_{\mathrm{p}}$ (see Figure \ref{fig:Kp}). \citealt{Yan2018} uses $K_{\mathrm{p}} = 268.7^{+6.2}_{-6.4}$ km s$^{-1}$ from their analysis of atmospheric dynamics using $\mathrm{H\alpha}$ as a tracer as well as $K_\star = 0.276 \pm 0.079$ km s$^{-1}$ from the discovery paper; however our measurement of $K_\star$ is nearly a factor of 1.2 ($\sim 0.42\sigma$ difference) less than that in the discovery paper. \citealt{Hoeijmakers2019} uses a similar value of $K_{\mathrm{p}}$ as ours that agrees within errors, which they obtained from fitting Fe II lines as well. Their procedure slightly differs from ours in regards to order of operations: instead of fitting their cross-correlated planet absorption track across all fully in-transit observations simultaneously for a given dataset, they fit each observation with an independent Gaussian absorption feature and fit a single circular orbital solution to the individual planet RVs they obtain across both of their datasets. However, they use the planet mass from \citealt{Gaudi2017} along with their value of $K_{\mathrm{p}}$ to revise the stellar mass, but then update the planet mass according to the value of $K_\star$ from \citealt{Gaudi2017} and their revised stellar mass. Thus their analysis is not self-consistent.  Our analysis, on the other hand, is self-consistent, because we obtain empirical values of $K_{\mathrm{p}}$ and $K_\star$ exclusively by applying the same routines and wavelength-consistent template linelists on all the data we present in this work.

\par Studies of KELT-9 b's neutral iron emission such as \citealt{Pino2020} and \citealt{Kasper2021} obtain complementary constraints on $K_{\mathrm{p}}$. In principle, one might expect these dayside measurements to disagree with our terminator measurements due to atmospheric circulation. However, the measurements of $K_{\mathrm{p}}$ from both \citealt{Pino2020} (displayed in Figure \ref{fig:Kp}) and \citealt{Kasper2021} (not included in Figure \ref{fig:Kp} since this work did not consolidate their multiple constraints from different nights of observations into one measurement) agree within $1\sigma$. This strengthens our claim that the planet's orbital parameters can be procured from the planet's transmission absorption signature without secondary effects from atmospheric processes like circulation or condensation.

\par We also find that our measurement of stellar mass agrees with \citealt{Gaudi2017} and \citealt{Hoeijmakers2019}, but disagrees with \citealt{Yan2018} by 1.1$\sigma$.

\par Since stellar density is a direct observable from transit lightcurves \citep{Seager2003}, we can use our revised stellar mass measurement and empirical measurements of stellar radius from literature to compute a stellar density and compare with the stellar density derived from the lightcurve as provided in \citealt{Gaudi2017}. The stellar radius provided in \citealt{Gaudi2017} is directly constrained to be $R_\star = 2.17 \pm 0.33\ \mathrm{R_\odot}$ from the Hipparcos parallax, effective temperature, bolometric flux of the star from integrating its spectral energy distribution (SED), and interstellar extinction. Using our measurement of the host star's dynamical mass and $V = \frac{4}{3} \pi R_\star^3$ (assuming a spherical star) with the \citealt{Gaudi2017} measurement for $R_\star$, we obtain a stellar density of $\rho_\star = 0.29\pm0.17\ \mathrm{g\ cm^{-3}}$, which is consistent with direct measurement from the lightcurve in \citealt{Gaudi2017} of $\rho_\star = 0.2702 \pm 0.0029\ \mathrm{g cm^{-3}}$. This suggests that our dynamical mass analysis agrees with independent metrics from previous studies. Note that the error on our empirical measurement of stellar density has nearly equivalent contributions from the error on the empirical radius ($\sim$55\% of the stellar density error) and our empirical mass error ($\sim$45\% of the stellar density error).

\par We note that the spherical star assumption is not necessarily valid for fast rotators. \citealt{Ahlers2020} goes an additional step to account for effects from gravity-darkening in their measurement of stellar radius using TESS transit lightcurves of KELT-9 b. Gravity-darkening is a phenomenon by which effective temperature varies across a stellar surface due to a rapidly rotating star's oblateness perturbing the star's hydrostatic equilibrium near its equator \citep{Ahlers2020}. While \citealt{Ahlers2020} do provide an equatorial radius and oblateness parameter that can, in principle, be combined with our mass measurement to recover a stellar density, we do not perform this step for comparison because the stellar equatorial radius they derive assumes a prior on the stellar mass based on \citep{Gaudi2017} of  $M_*=2.52^{+0.25}_{-0.20}~M_\odot$, which differs from our measurement of $M_*=2.11\pm0.78\ \mathrm{M_\odot}$.

\section{Discussion}
\label{sec:discussion}

\subsection{Caveats to the orbital motion model}
The possibility of an eccentric orbit would undermine our measurement of $K_{\mathrm{p}}$ and by extension the KELT-9 b's mass. Our model presumes that KELT-9 b and its host star obey circular orbital motion based on the planet's estimated circularization timescale \citealt{Borsa2019}. Furthermore, we obtain an empirical constraint of $e\cos{\omega} = 4.77 \times 10^{-7} \pm 7.2 \times 10^{-5}$ from the ephemerides of the primary transit and secondary eclipse; the precision of this constraint is possible due to the inclusion of TESS observations in our global fit for the system ephemerides. With such a low value for $e\cos{\omega}$, it is unlikely that the planet's orbit is significantly eccentric enough to affect our empirical measurement of the planet's mass within errors.

\par We also assume that the dominant effect of atmospheric dynamics on the planet's absorption signature between 2nd and 3rd contact is a constant RV offset due to day-to-nightside winds. \citealt{Ehrenreich2020} has shown that the wind component of WASP-76 b's atmospheric absorption signature changes from 0 km s$^{-1}$ to $\sim$11 km s$^{-1}$ over the course of the planet's transit. This would imply that our model of the planet's atmospheric absorption signature may not sufficiently capture the effect of KELT-9 b's atmospheric dynamics. However, we note that unlike \citealt{Ehrenreich2020}, our analysis does not include observations taken during ingress or egress in the procedure for fitting the orbital motion of the planet. Ingress and egress are expected to contribute the strongest asymmetries in the absorption signal due to eastward equatorial jets and rotation, yielding an additional redshift on the leading limb during ingress and a blueshift on the trailing limb during egress (assuming the planet's rotational axis and orbital axis are aligned); \citealt{Ehrenreich2020} demonstrated this effect empirically. Since we exclude observations that capture a partial limb of the planet, these asymmetries should not manifest as strongly in our analysis. 

\par Even if we were to include the entire transit in our analysis, the absorption signal during ingress and egress is not very strong in our observations. When we align all in-transit observations between 1st and 4th contact in the planet's restframe and additionally remove the measured offset component for each dataset, the absorption feature is visibly centered around 0 km s$^{-1}$ over the course of the entire transit. This suggests that the wind speed does not appear to vary significantly over the course of the transit.  Furthermore, the asymmetry observed by \citealt{Ehrenreich2020} shows a constant RV offset during the second half of the transit, which is where KELT-9 b's absorption signal is strongest and will contribute the most to our measured day-to-nightside wind measurement.

\subsection{Complementary constraints on system parameters}
\label{sec:overconstrained}
Among its numerous significant attributes, one noteworthy characteristic of the KELT-9 system is that its physical properties are empirically over-constrained.  In particular, additional empirical constraints are possible due to the gravity-darkening signature during the primary transit that originates from rotational flattening of the host star. The geometry of such systems introduces mathematical complexities that make quantitative analyses analytically intractable. Rather, numerical methods are required to obtain the system parameters; these are beyond the scope of this work. Therefore, we proceed to qualitatively describe the ways in which a complete, model-independent solutions to the system parameters can be achieved.

\par The first method treats the system as a double-lined eclipsing binary (SB2) as done in this work. We have determined the masses of the system from spectroscopic observables in this work by treating the KELT-9 system as an eclipsing double-lined spectroscopic binary; thus all the orbital elements ($K_\star$, $K_{\mathrm{p}}$, $a_\star$, $a_{\mathrm{p}}$, etc.) of the system are known. Furthermore, one can apply the assumption that the orbit is circular to get the stellar and planetary semi-major axes $a_\star$ and $a_{\mathrm{p}}$ respectively.  Determining the radii of the host star and planet is complicated somewhat by the fact the star is oblate and gravity-darkened.  However, the photometric gravity-darkening signature during the primary transit (as presented in \citealt{Ahlers2020}) uniquely constrains the inclination of the stellar rotational axis as well as the stellar oblateness parameter. These two parameters then completely specify the geometry of the planet's transit across the oblate stellar disk, taking into account the planet's orbital inclination (angular offset along the line-of-sight between the plane of the planet's orbit and the plane of the sky) and stellar inclination (angular offset along the line-of-sight between the stellar rotation axis and the plane of the sky), and spin-orbit obliquity (angular offset in the plane-of-sky between the stellar rotation axis and the planet's orbital axis). The chord transited by the planet is related to the equatorial stellar radius by these angular quantities and the oblateness of the star. From this, the equatorial stellar radius can be related to the orbital semi-major axis by the FWHM transit duration and orbital period. The planet's radius is consequently obtained from the ratio of planet radius to stellar equatorial radius provided in \citealt{Ahlers2020}. Thus a  complete mass-radius solution to the KELT-9 system can be obtained. 

\par A second approach is to use just the host star RVs in combination with an empirical stellar radius from an SED and parallax. The latter part of this approach is provided in \citealt{Gaudi2007} using KELT-9's Hipparcos parallax, yielding $R_\star = 2.17 \pm 0.33\ \mathrm{R_\odot}$. This empirical radius is more appropriately thought of as an effective radius since it assumes a spherical star with a uniformly illuminated sky-projected disk. This empirical radius can be translated into a measurement of the true stellar equatorial radius by determining the equatorial radius (given the oblateness parameter from \citealt{Ahlers2016}) that yields an equivalent integrated disk flux as the spherical case. Numerically integrating the oblate star disk flux as done in \citealt{Ahlers2016} requires accounting for: 1) the star's longitudinal variation in flux from limb-darkening (Equation 3 of \citealt{Ahlers2016}), 2) latitudinal variation in flux from gravity-darkening (Equation 2 of \citealt{Ahlers2016}), and 3) inclination of the stellar rotational axis, which affects both the sky-projected disk shape of the oblate star and the range of latitudes visible along the line-of-sight (relevant for the latitude-dependence of the gravity-darkening profile). As before, the planet's radius is deduced from the ratio of planet radius to stellar equatorial radius given in \citealt{Ahlers2020}. The stellar mass can be determined from the stellar radial parameters in combination with stellar density, a transit observable as given in \citealt{Seager2003}. The relation for stellar density will require modifications for the geometry of an oblate and inclined spheroid star transited by a planet on an inclined and oblique circular orbit. These nuances directly effect the relations between transit observables, i.e. depth and duration, and the system parameters of interest, i.e. $a/R_{\star, \mathrm{eq}}$ and $\rho_\star$. Lastly, the planet mass can be related to the stellar mass by the stellar RV semi-amplitude as given in Equation 9 of \citealt{Hoeijmakers2019} for the complete solution to the system. A revised measurement of the empirical stellar radius with a Gaia parallax would significantly improve the precision of this constraint. 

\par The third approach combines the single-lined spectroscopy measurement of $K_\star$ with the photometric gravity-darkening signature during primary transit, which provides constraints on stellar surface gravity. Since the host star is an oblate spheroid, the surface gravity varies spatially as given by Equation 10 of \citealt{Barnes2009}. The transit signature maps stellar surface brightness as the planet crosses the stellar disk. For a given $R_\star$ and external constraint on $v\sin{i_\star}$, one can uniquely determine limb-darkening coefficients, oblateness, stellar inclination, orbital inclination, and spin-orbit obliquity as done in \citealt{Ahlers2020}. For a given stellar equatorial radius, the system can be uniquely solved. Stellar mass is related to stellar radius using the stellar density, a transit observable (taking into account the nuances in the case of an oblate star as described above). Then as before, planet radius can be obtained from the planet-to-star radius ratio from the gravity-darkening transit signature and the planet mass comes from Equation 9 of \citealt{Hoeijmakers2019}. The only missing component to solve the system is the stellar radius. \citealt{Ahlers2020} uses a gravity-darkening correction on the SED from \citealt{Gaudi2017}, similar to our description in the second approach. To distinguish this third approach from the second, we note that stellar equatorial radius is related to the gravity-darkening exponent $\beta$ and stellar inclination by plugging in Equation 10 of \citealt{Barnes2009} into Equation 9 of the same work. These quantities are not degenerate because their effects are not scaled versions of each other. Thus, it is possible to determine a unique solution for the equatorial radius based on the shape of the transit lightcurve, and thus all the other geometric parameters parameters (stellar oblateness, inclinations, spin-orbit obliquity, limb-darkening coefficients, and the gravity-darkening exponent).

\par The fourth and final approach we present combines the single-lined spectroscopy measurement of $K_\star$ with stellar surface gravity as measured by the broadening of the stellar absorption lines seen in high-resolution spectra. This measurement of $\log(g_\star)$ does not account for the fact that surface gravity varies spatially across the surface of an oblate star. Instead it serves more as an effective surface gravity assuming a spherical star. Under this assumption, Section 2.3.1 of \citealt{Stevens2018} presents a derivation of system parameters from spectroscopic stellar RV semi-amplitude and spectroscopic stellar surface gravity. This constraint is the weakest because it does not account for the oblate geometry of the star. Additionally, spectroscopically determining the surface gravity of hot, rapid rotators is challenging since their absorption features are dominated by rotational broadening.

\par We hope these four approaches guide future follow-up work to cross-validate and tighten constraints on the KELT-9 system and other transiting two-body systems in which the primary is a rapid rotator.

\subsection{Independent metrics of mass in the literature}
With sufficient photometric precision, the BEaming, Ellipsoidal, and Reflection (BEER) algorithm \citep{Faigler2011} enables the detection of short-period massive planets, both transiting and non-transiting. This technique encompasses three distinct photometric signatures, of which two (Doppler beaming and ellipsoidal variations) depend on the mass of the planet. We provide order-of-magnitude estimates of these signals based on our revised mass constraints.

\par Doppler beaming is the periodic variation in flux due to the orbital motion of a star hosting a companion \citep{Loeb2003}. The fractional amplitude in flux modulations is given in \citealt{Loeb2003} as 
\begin{equation}
\frac{\Delta F}{F_0} = \frac{3 - \alpha_{\mathrm{beam}}}{c}  K_\star
\end{equation}
where $\alpha_{\mathrm{beam}}$ is the power law exponent of the emitted flux from the source star as a function of frequency. We can estimate $\alpha_{\mathrm{beam}}$ of a blackbody source with effective temperature $T_{\mathrm{eff}}$ according to Equation 3 in \citealt{Loeb2003} as $\alpha_{\mathrm{beam}}(\nu) \approx \frac{\mathrm{e}^x (3-x) - 3}{\mathrm{e}^x - 1}$ where $x = \frac{h \nu}{k T_{\mathrm{eff}}}$. Adopting the frequency corresponding to the $V$-band wavelength of 551 nm as a representative frequency, we obtain a Doppler beaming signal on the order $\frac{\Delta F}{F_0} = 2.17$ ppm. This estimated signal falls far below the photometric precision of present-day instruments like TESS to be discernable.


\par Ellipsoidal variations are flux perturbations that arise as a consequence of tidal effects induced by a companion. Equation 2 of \citealt{Faigler2011} estimate ellipsoidal variations of amplitudes
\begin{equation}
    A_{\mathrm{ellip}} = \alpha_{\mathrm{ellip}} \frac{m_{\mathrm{p}} \sin{i_{\mathrm{orbit}}}}{M_\star} \Big(\frac{R_\star}{a}\Big)^3 \sin{i_{\mathrm{orbit}}}
\end{equation}
where $\alpha_{\mathrm{ellip}}$ is defined in Equation 4 of \citealt{Faigler2011} as
\begin{equation}
    \alpha_{\mathrm{ellip}} \approx 0.15\frac{(15+u)(1+g)}{3-u}
\end{equation}
The parameter $u$ is the linear limb-darkening coefficient of the star, which we take to be $u=0.3356$ based on \citealt{Claret2017}. We adopt edge cases of $g = 0.3$ and $g = 1$ for the gravity-darkening coefficient (not to be mistaken with the gravity-darkening exponent from \citealt{Ahlers2020}) as suggested in \citealt{Faigler2011}. The ratio $\frac{R_\star}{a}$ is a direct observable from transit lightcurves; we use $\frac{a}{R_\star}=3.153$ from \citealt{Gaudi2017}. The resulting estimate of the ellipsoidal variations signal based on our mass ratio ranges between 35.0 and 53.8 ppm.  

\par \citealt{Wong2020} presents photometric harmonics in the phase curve of the KELT-9 system by phase-folding TESS lightcurves and binning at 30-minute intervals. They measure periodic variations attributable to ellipsoidal variations with an amplitude of $39.6 \pm 4.5$ ppm. This measurement is within the range of the ellipsoidal variation signal amplitudes we estimate based on our revised masses of the system. In fact, it is marginally easier to explain \citealt{Wong2020}'s measurement with our planet-star mass ratio than with the original value in \citealt{Gaudi2017}, which yield ellipsoidal variation signal amplitudes between 38.9 to 60.0 ppm for the same range of values for $\beta$. Note that numerous uncertainties play into our estimate of ellipsoidal variations, such as the uncertainty of the limb- and gravity-darkening coefficients of this particular star. In particular, metallicity and evolutionary effects on the limb-darkening coefficient are stronger after the onset of convection ($\log{T_{\mathrm{eff}}} > 3.9$) \citep{Claret2017}. Furthermore, the star is significantly distorted by rotation and the planet's orbit is nearly orthogonal to this distortion. This geometry limits gravitational distortion of the star by the companion planet and weakens the signal from ellipsoidal variations.

\subsection{Consequences for atmospheric escape}


As the hottest planet known to-date, KELT-9 b is a unique target for studying the hydrodynamic escape of planetary atmospheres driven by extreme stellar irradiation \citep{Fossati2018, GarciaMunoz2019, Krenn2021}. Observations of atmospheric loss on KELT-9 b use light element tracers such as Balmer lines \citep{Yan2018, Cauley2019, Wyttenbach2020}  and Mg I \citep{Cauley2019} to probe mass loss rates according to empirically-validated species densities and a slew of assumptions surrounding elemental abundances and equilibrium conditions. Notably, these studies generally assume thermodynamic and ionization equilibrium as well as solar abundances of species; the former may be discounted by NLTE studies of KELT-9 b's upper atmosphere \citep{GarciaMunoz2019, Wyttenbach2020, Fossati2021}, while the latter may be invalidated by recent work on atmospheric retrieval of hot Jupiters \citep{Giacobbe2021}. 

\par We summarize the state of the field to-date regarding the mass loss rate of KELT-9b. \citealt{Gaudi2017} provided an initial estimate of KELT-9 b's mass loss rate between $10^{10}$ -- $10^{13}$ g s$^{-1}$ based on the equation for energy-limited mass loss rate provided in Equation 22 of \citealt{Murray-Clay2009}, which scales inversely with the mass of the planet. \citealt{Yan2018} empirically constrain KELT-9 b's mass loss rate by identifying an excess in H$\alpha$ absorption depth relative to the photometric transit depth in the continuum. By estimating 1) the number density of hydrogen from model fits to the H$\alpha$ line profile (these models depend on planetary parameters, most notably planet mass) and 2) the contribution from the altitude regime of the planet where H$\alpha$ can energetically escape beyond the planet's Roche lobe, \citealt{Yan2018} estimate a mass loss rate of $\dot{M} \sim 10^{12}$ g s$^{-1}$. \citealt{Cauley2019} corroborates this measurement within an order of magnitude, estimating $\dot{M} \sim 1 \times 10^{12}$ g s$^{-1}$ when using Mg I as a tracer and $\dot{M} \sim 3 \times 10^{12}$ g s$^{-1}$ from Balmer line analysis. \citealt{Wyttenbach2020} report $\dot{M} \sim 10^{12.8 \pm 0.3}$ g s$^{-1}$ from Balmer line analysis as well.

\par Reconciling observations of KELT-9 b's atmospheric escape with theory currently faces unresolved challenges. HJ atmospheric escape is typically modelled as hydrodynamic escape due to heating from Lyman continuum absorption in the X-ray and extreme-ultraviolet (XUV). \citealt{Fossati2018} notes that ionizing XUV fluxes in the wavelength regime relevant to heating (and thus escape) are weaker in hotter intermediate mass stars, such as the stellar host of the KELT-9 system, than their cooler ($\sim$8000-8500 K) counterparts, such as the host of another UHJ WASP-33 b. XUV flux is driven by coronal heating, which is related to stellar magnetic activity. A convective envelope is necessary for the interactions that generate magnetic activity. The surface convective zone vanishes in hotter stars and consequently they emit less XUV flux. \citealt{Fossati2018} estimate KELT-9 b's mass loss rate is $\sim$10$^{10}$ - 10$^{11}$ by accounting for the heating efficiency of XUV flux from the KELT-9 host star. Their estimated H$\mathrm{\alpha}$ transit depth of 0.7\% does not agree with the observed 1.8\% in \citealt{Yan2018}. They propose that one way of bridging this gap is by adopting a planetary mass on the lower end of the 1$\sigma$ range provided in \citealt{Gaudi2017} ($M_{\mathrm{p}} = 2.88 \pm 0.84\ \mathrm{M_J}$). As previously noted, $\dot{M}$ scales inversely with planet mass in the most simplified energy-limited case, which balances gravitational potential and thermal heating from the host star's X-ray and extreme UV (XUV) radiation. This would bring the planetary mass required to explain observations of KELT-9 b's atmospheric loss in closer agreement with our revised empirical mass. We estimate that our revised mass increases the estimated energy-limited mass loss rate (see \citealt{Fossati2018} for a discussion of this estimate for KELT-9 b and its nuances) by $\sim$33\%. Note that recent work by \citealt{Krenn2021} comparing the energy-limited approximation against hydrodynamic simulations of atmospheric escape shows that the energy-limited approximation is not suited for estimating UHJ mass-loss rates (or planets in extreme temperature or mass regimes in general; UHJs satisfy both of these conditions).

\par \citealt{GarciaMunoz2019} expands upon the implications of \citealt{Fossati2018} by proposing that Balmer continuum absorption in the near-ultraviolet (NUV) is the dominant source of heating. However, their models best match observations of KELT-9 b's H$\alpha$ absorption when adopting a planet mass between 0.80--1.20 $M_J$. This is incompatible with our revised mass. 

\section{Conclusion} \label{sec:conclusion}
\par We have presented the analysis of spectroscopic data of KELT-9b in the context of an eclipsing double-lined spectroscopic binary.  This has enabled us to direcly and empirically obtain the dynamical mass the KELT-9 b and its host star. Using multi-epoch spectroscopic observations of the system, we find that the dynamical mass of the planet is $m_{\mathrm{p}} = 2.17 \pm 0.56\ \mathrm{M_J}$ and the star is $M_\star = 2.11 \pm 0.78\ \mathrm{M_\odot}$. Our planet mass measurement generally agrees with previous literature \citep{Gaudi2017, Hoeijmakers2019}. We also obtain a purely empirical measurement of stellar density, a direct observable from transit lightcurves, that agrees with the value in the discovery paper; this suggests that our analysis is trustworthy.  Our methodology can be applied to many Hot Jupiter systems, thereby enabling the direct and empirical measurement of their planets and host stars.

\par We note that the KELT-9 system is empirically overconstrained due to the unique geometric information provided by the in-transit gravity-darkening signature of rapid rotator systems. We present a framework for obtaining a complete solution to the system parameters in three purely observational ways.

\par An order-of-magnitude estimate shows that this revised planetary mass is large enough to induce ellipsoidal variations observable with TESS phase curves of the KELT-9 system. Furthermore, this result is especially crucial for studies of atmospheric escape, which depend on mass for empirical measurement and theoretical modelling; as of now, the high levels of atmospheric escape seen on KELT-9 b have not been reconciled with its high mass. Our purely empirical confirmation of the planet's mass motivates further exploration of this conundrum surrounding the substantial mass loss on KELT-9 b.

\acknowledgments
We thank the anonymous referee for taking the time to provide insightful comments which improved the quality of this paper. A.P.A would like to thank the David G. Price Fellowship in Astronomical Instrumentation for funding her work this year. Work by B.S.G. and J.W. was partially supported by the Thomas Jefferson
Chair for Space Exploration endowment from the Ohio State University.  This work is based on observations made with the Large Binocular Telescope. The LBT is an international collaboration among institutions in the United States, Italy and Germany. LBT Corporation partners are: The University of Arizona on behalf of the Arizona Board of Regents; Istituto Nazionale di Astrofisica, Italy; LBT Beteiligungsgesellschaft, Germany, representing the Max-Planck Society, The Leibniz Institute for Astrophysics Potsdam, and Heidelberg University; The Ohio State University, representing OSU, University of Notre Dame, University of Minnesota and University of Virginia. This paper includes data collected by the TESS mission. Funding for the TESS mission is provided by the NASA's Science Mission Directorate. We thank Dr. George Zhou for providing TRES observations of the KELT-9 system. We also thank Dr. Marshall Johnson, Dr. Francesco Borsa, Dr. Jens Hoeijmakers, Dr. Lorenzo Pino, and Dr. Luca Fossati for contributing their valuable expertise regarding high-resolution spectroscopy and the KELT-9 system. \\ \\

\facilities{LBT (PEPSI), TNG (HARPS-N), Fred L. Whipple Observatory (TRES)} \\
\software{scipy \citep{scipy2020}, petitRADTRANS \citep{Molliere2019}, SME \citep{Valenti1996, Valenti2012}, Time Utilities \citep{Eastman2012}, emcee \citep{Foreman-Mackey2013}, george (\texttt{https://george.readthedocs.io/en/latest/})}

\clearpage
\bibliographystyle{aasjournal}
\bibliography{references}

\appendix
\section{Error propagation of stellar and planetary mass from RV observables}
\label{sec:errorProp}
We apply linear propagation of errors to recover the following expression for the propagated uncertainty of the planet mass:
\begin{equation*}
    \label{eq:m_p_err}
    \begin{split}
    \delta m_{\mathrm{p}} =& \Bigg[\Big(\frac{\partial m_{\mathrm{p}}}{\partial P}\Big)^2 \delta P^2 + \Big(\frac{\partial m_{\mathrm{p}}}{\partial i}\Big)^2 \delta i^2 + \Big(\frac{\partial m_{\mathrm{p}}}{\partial K_{\mathrm{p}}}\Big)^2 \delta K_{\mathrm{p}}^2
    + \Big(\frac{\partial m_{\mathrm{p}}}{\partial K_\star}\Big)^2 \delta K_\star^2\Bigg]^{1/2}
    \end{split}
\end{equation*}
We present the analytic form of each partial derivative term in Equation \ref{eq:m_p_err} below:
\begin{align*}
    \label{eq:m_p_partials}
    \frac{\partial m_{\mathrm{p}}}{\partial P} &= \frac{K_\star (K_{\mathrm{p}}+K_\star)^2 }{2\pi G \sin^3{i}}\\
    \frac{\partial m_{\mathrm{p}}}{\partial i} &= \frac{-3 K_\star (K_{\mathrm{p}}+K_\star)^2 P \cos{i}}{2\pi G \sin^2{i}}\\
    \frac{\partial m_{\mathrm{p}}}{\partial K_{\mathrm{p}}} &= \frac{K_\star (K_{\mathrm{p}}+K_\star) P}{\pi G \sin^3{i}}\\
    \frac{\partial m_{\mathrm{p}}}{\partial K_\star} &= \frac{(K_{\mathrm{p}}+K_\star)(K_{\mathrm{p}}+3K_\star) P}{2\pi G \sin^3{i}}
\end{align*}
Analogously, the propagated stellar mass uncertainty is:
\begin{equation*}
    \label{eq:M_star_err}
    \begin{split}
    \delta M_\star =& \Bigg[\Big(\frac{\partial M_\star}{\partial P}\Big)^2 \delta P^2 + \Big(\frac{\partial M_\star}{\partial i}\Big)^2 \delta i^2 + \Big(\frac{\partial M_\star}{\partial K_{\mathrm{p}}}\Big)^2 \delta K_{\mathrm{p}}^2
    + \Big(\frac{\partial M_\star}{\partial K_\star}\Big)^2 \delta K_\star^2\Bigg]^{1/2}
    \end{split}
\end{equation*}
with partial derivative terms of the form
\begin{align*}
    \label{eq:M_star_partials}
    \frac{\partial m_{\mathrm{p}}}{\partial P} &= \frac{K_{\mathrm{p}} (K_{\mathrm{p}}+K_\star)^2 }{2\pi G \sin^3{i}}\\
    \frac{\partial m_{\mathrm{p}}}{\partial i} &= \frac{-3 K_{\mathrm{p}} (K_{\mathrm{p}}+K_\star)^2 P \cos{i}}{2\pi G \sin^2{i}}\\
    \frac{\partial m_{\mathrm{p}}}{\partial K_{\mathrm{p}}} &= \frac{(K_{\mathrm{p}}+K_\star)(3K_{\mathrm{p}}+K_\star) P}{2\pi G \sin^3{i}}\\
    \frac{\partial m_{\mathrm{p}}}{\partial K_\star} &= \frac{K_{\mathrm{p}} (K_{\mathrm{p}}+K_\star) P}{\pi G \sin^3{i}}
\end{align*}

\end{CJK*}

\end{document}